\documentclass[preprint2]{aastex}

\usepackage{natbib}
\usepackage{graphicx}
\usepackage{amssymb}
\usepackage{epstopdf}
\usepackage[xindy, toc, hyperfirst=false, nolist, nostyles, sanitize={name=false,description=false,symbol=false}]{glossaries}
\glsdisablehyper

\newcommand{\msun}{\hbox{$M_{\odot}$}}

\newglossaryentry{vrad}{name={radial velocity~}, text={radial velocity}, symbol={\ensuremath{v_\textrm{rad}}}, description={radial velocity}, sort=vrad}
\newglossaryentry{vrot}{name={stellar rotation~}, name={stellar rotation}, symbol={\ensuremath{v_\textrm{rot}}}, description={radial velocity}, sort=vrot}
\newcommand{\vrot}{\glssymbol*{vrot}}
\newcommand{\vrad}{\glssymbol*{vrad}}

\newcommand{\rasc}[4]{\hbox{\ensuremath{\alpha=#1^h#2^m#3\overset{s}{.}#4}}}
\newcommand{\decl}[3]{\hbox{\ensuremath{\delta=#1^{\circ}#2\arcmin#3\arcsec}}}
\newcommand{\kpc}{kpc}

\newcommand{\kms}{\ensuremath{\textrm{km}~\textrm{s}^{-1}}}

\newcommand{\xray}{X-ray}

\newcommand{\masyr}{\ensuremath{\textrm{mas~yr}^{-1}}}

\newglossaryentry{angstrom}{name=\AA, description={unit of length $10^{-10}$\,m}, sort=angstrom}
\newglossaryentry{nir}{name=NIR,description={near infrared},first = {near infrared (NIR)}}
\newglossaryentry{psf}{name=PSF,description={Point Spread Function},first = {PSF}}
\newglossaryentry{fwhm}{name=FWHM,description={Full Width Half Maximum},first = {FWHM}}
\newglossaryentry{rms}{name=RMS,description={Root Mean Square},first = {RMS}}
\newglossaryentry{uv}{name=UV,description={ultra violet},first = {ultra violet (UV)}}
\newglossaryentry{halpha}{name=\ensuremath{\textrm{H}\alpha}, description={First line of the Balmer series at 6563\,\AA}, sort=halpha}
\newglossaryentry{mgb}{name={Mg \textsc{i} b}, description={Triplet at 5167\,\AA, 5173\,\AA and 5184\,\AA}}
\newglossaryentry{sobolevapprox}{name={Sobolev approximation}, description={Lines are approximation with an infinitley thin interaction region \citep[e.g. no broadening][]{1960mes..book.....S}}, first={Sobolev approximation }}
\newglossaryentry{radeq}{name={radiative equilibrium}, description={The net flux of energy between matter and radiation field is zero}}
\newglossaryentry{nebularapprox}{name={nebular approximation}, description={Assumes that the plasma condition are controlled by a central radiation source. The radiation field decreases with the distance to the source by geometrical dilution. See \citet{1978stat.book.....M} for details}}
\newglossaryentry{modnebularapprox}{name={modified nebular approximation}, description={In contrast to \gls{nebularapprox} where only geometrical dilution is taken into account, the modified nebular approximation also takes dilution by other radiative processes into account }, first={modified nebular approximation}, parent=nebularapprox}
\newglossaryentry{thompsonscat}{name={Thomson scattering}, description={Scattering of photons on low energy electrons}}
\newglossaryentry{lte}{name={LTE}, description={Local Thermodynamic Equilibrium}, first={local thermodynamic equilibrium (LTE)}}
\newglossaryentry{lsr}{name={LSR}, description={Local Standard of Rest}, first={\textit{local standard of rest} (LSR)}}
\newglossaryentry{mc}{name={MC}, description={Monte Carlo}, first={\textit{Monte Carlo} (MC)}}

\newglossaryentry{sfit}{name=SFIT, text=\textsc{sfit}, description={spectral fitting program for hot stars \citep{2001A&A...376..497J}}, first={\textsc{sfit} \citep{2001A&A...376..497J}}}
\newglossaryentry{iraf}{name=IRAF, text=\textsc{iraf}, description={Image Reduction and Analysis Facility maintained by NOAO}, first={\textsc{iraf}\protect\footnote{IRAF: the Image Reduction and Analysis Facility is distributed by the National Optical Astronomy Observatory, which is operated by the Association of Universities for Research in Astronomy (AURA) under cooperative agreement with the National Science Foundation (NSF).}}}
\newglossaryentry{pyraf}{name=PyRAF, text=\textsc{PyRAF}, description={Python wrap of \gls{iraf} maintained by STSCI}, first=\textsc{PyRAF} \protect\footnote{PyRAF is a product of the Space Telescope Science Institute, which is operated by AURA for NASA.}}
\newglossaryentry{scipy}{name=SciPy, text=\textsc{Scipy}, description={Scientific Python \cite{Jones:2001fk}}}
\newglossaryentry{moog}{name=MOOG,text={\textsc{moog}}, description={spectral synthesis software \citep{1973ApJ...184..839S}}, first={\textsc{Moog} \citep{1973ApJ...184..839S}}}
\newglossaryentry{atlas9}{name=ATLAS9,description={grid of stellar atmospheres \citep{2004astro.ph..5087C}}, first={ATLAS9 \citep{2004astro.ph..5087C}}}
\newglossaryentry{vald}{name=VALD,description={Vienna Atomic Line Database \citep{2000BaltA...9..590K}}, first={Vienna Atomic Line Database \citep[VALD;][]{2000BaltA...9..590K}}}
\newglossaryentry{sextractor}{name=SExtractor, text=\textsc{SExtractor}, description={Source Extractor photometry program \citep{1996A&AS..117..393B}}, first={\textsc{SExtractor} \citep{1996A&AS..117..393B}}}
\newglossaryentry{idl}{name=IDL,text={\textsc{idl}}, description={Interactive Data Language}}
\newglossaryentry{makee}{name=MAKEE,text=\textsc{makee}, description={MAuna Kea Echelle Extraction by Tom Barlow available}}
\newglossaryentry{minuit}{name=MINUIT,text={\textsc{minuit}}, description={collection of numerical optimization tools \citep{James:1975dr}}}
\newglossaryentry{migrad}{name=MIGRAD,text={\textsc{migrad}}, description={numerical gradient optimization tools - part of \gls{minuit}}}

\newglossaryentry{2mass}{name=2MASS,description={Two Micron All Sky Survey \citep{2006AJ....131.1163S}}}

\newglossaryentry{scp}{name=SCP,description={Supernova Cosmology Project, led by Saul Perlmutter}, first={Supernova Cosmology Project (SCP)}}
\newglossaryentry{hzsns}{name=HZSNS,description={High Z Supernova Search, led by Brian Schmidt}, first={High Z Supernova Search (HZSNS)}}
\newglossaryentry{vlt}{name=VLT,description={Very Large Telescope located on Cerro Paranal (Chile)}, first={Very Large Telescope (VLT)}}
\newglossaryentry{flames}{name=FLAMES,description={Multi-object, intermediate and high resolution spectrograph mounted on the  \gls{vlt}}}
\newglossaryentry{hires}{name=HIRES, description={High Resolution Echelle Spectrometer mounted on the Keck Telescope}, first={High Resolution Echelle Spectrometer \citep[HIRES;][]{1994SPIE.2198..362V}}}
\newglossaryentry{lris}{name=LRIS,description={Low Resolution Imaging Spectrometer mounted on the Keck Telescope}, first={Low-Resolution Imaging Spectrometer \citep[LRIS;][]{Oke95}}}

\newglossaryentry{essence}{name=ESSENCE,description={The `Equation of State: SupErNovae trace Cosmic Expansion' project \citep[ESSENCE;][]{2002AAS...201.7809G}}, first={`The Equation of State: SupErNovae trace Cosmic Expansion' \citep[ESSENCE;][]{2002AAS...201.7809G}}}
\newglossaryentry{ifu}{name=IFU,description={Optical instrument combining spectrographic and imaging capabilities, used to obtain spatially resolved spectra}, first={Integral Field Unit (IFU)}, firstplural={Integral Field Units (IFUs)}} 

\newglossaryentry{besancon}{name=Besan\c{c}on Model, description={Model of stellar population synthesis of the Galaxy, including kinematics.}}

\newglossaryentry{int}{name=INT,description={Isaac Newton 2.5\,m Telescope}, first={Isaac Newton 2.5\,m Telescope (INT)}}
\newglossaryentry{iau}{name=IAU,description={International Astronomical Union}, first={IAU}}
\newglossaryentry{chandra}{name=Chandra,description={Chandra \xray\ Observatory (space-based)}}
\newglossaryentry{hst}{name=HST,description={Hubble Space Telescope}}
\newglossaryentry{wfpc2}{name=WFPC2,description={Wide-Field Planetary Camera 2 mounted on the \gls{hst}}, first={Wide-Field Planetary Camera 2 (WFPC2)}}
\newglossaryentry{acs}{name=ACS,description={Advanced Camera for Surveys mounted on the \gls{hst}}, first={Advanced Camera for Surveys (ACS)}}
\newglossaryentry{snls}{name=SNLS,description={Supernova Legacy Survey \citep{2003AAS...203.8209P}}, first={Supernova Legacy Survey \citep[SNLS;][]{2003AAS...203.8209P}}}
\newglossaryentry{dass}{name=DASS, description={Digitized Astronomy Supernova Survey \citep{1975PASP...87..565C}}, first={Digitized Astronomy Supernova Survey \citep[DASS;][]{1975PASP...87..565C}}}
\newglossaryentry{bait}{name=BAIT, description={Berkley Automatic Imaging Telescope \citep{1993PASP..105.1164R}}, first={Berkley Automatic Imaging Telescope \citep[BAIT;][]{1993PASP..105.1164R}}}
\newglossaryentry{kait}{name=KAIT, description={Katzman Automatic Imaging Telescope \citep{2001ASPC..246..121F}}, first={Katzman Automatic Imaging Telescope \citep[KAIT;][]{2001ASPC..246..121F}}}
\newglossaryentry{loss}{name=LOSS, description={Lick Observatory Supernova Search  \citep{2000AIPC..522..103L}}, first={Lick Observatory Supernova Search \citep[LOSS;][]{2000AIPC..522..103L}}}
\newglossaryentry{ctss}{name=CTSS,description={Cal\'{a}n/Tololo Supernova Survey \citep{1993AJ....106.2392H}}, first={Cal\'{a}n/Tololo supernova survey \citep[CTSS;][]{1993AJ....106.2392H}}}
\newglossaryentry{ctio}{name= CTIO, description={Cerro Tololo Inter-American Observatory}, first={Cerro Tololo Inter-American Observatory (CTIO)}}
\newglossaryentry{ptf}{name=PTF, description={Palomar Transient Factory \citep{2009PASP..121.1334R}}, first={Palomar Transient Factory \cite[PTF;][]{2009PASP..121.1334R}}}
\newglossaryentry{batse}{name=BATSE, description={Burst and Transient Source Experiment mounted on the Compton Gamma Ray Observatory}, first={Burst and Transient Source Experiment (BATSE)}}
\newglossaryentry{bepposax}{name=BeppoSAX, description={\xray\ satellite named in honor of Giuseppe "Beppo" Occhialini}}
\newglossaryentry{rosat}{name=ROSAT, description={short for R\"{o}ntgensatellit}, first={ROSAT}}
\newglossaryentry{hete2}{name=HETE2, description={High Energy Transient Explorer}, first={High Energy Transient Explorer (HETE)}}
\newglossaryentry{gnirs}{name=GNIRS, description={Gemini Near InfraRed Spectrograph mounted on the Gemini North Telescope}}
\newglossaryentry{wifes}{name=WiFeS, description={Wide Field Spectrograph - \gls{ifu} mounted on the 2.3\,m telescope at Siding Spring Observatory}}
\newglossaryentry{swift}{name=Swift, description={Swift Gamma-Ray Burst Mission}}
\newglossaryentry{vla}{name=VLA, description={Very Large Array radio telescope located in North America}, first={Very Large Array (VLA)}}
\newglossaryentry{evla}{name=EVLA, description={Extended Very Large Array radio telescope located in North America}, first={Extended Very Large Array (EVLA)}}
\newglossaryentry{sdss}{name=SDSS, description={Sloan Digital Sky Survey}}
\newglossaryentry{dss}{name=DSS, description={Digitized Sky Survey}}
\newglossaryentry{skymapper}{name=SkyMapper, description={SkyMapper telescope \citep{2007PASA...24....1K}}, first={SkyMapper \citep{2007PASA...24....1K}}}
\newglossaryentry{panstarrs}{name=PanSTARRS, description={Panoramic Survey Telescope \& Rapid Response System \citep{2004SPIE.5489...11K}}, first={Panoramic Survey Telescope \& Rapid Response System \citep[PanSTARRS;][]{2004SPIE.5489...11K}}}
\newglossaryentry{lsst}{name=LSST, description={Large Synoptic Survey Telescope}, first={Large Synoptic Survey Telescope \citep[LSST;][]{2006AAS...209.8604P}}}
\newglossaryentry{ppmxl}{name=PPMXL, description={PPMXL Catalog of Positions and Proper Motions on the ICRS \citep{2010AJ....139.2440R}}}
\newglossaryentry{gaia}{name=GAIA, description={Global Astrometric Interferometer for Astrophysics \citep{2001A&A...369..339P}}, first={Global Astrometric Interferometer for Astrophysics \citep[GAIA;][]{2001A&A...369..339P}}}
\newglossaryentry{ligo}{name=LIGO, description={Laser Interferometer Gravitational Wave Observatory}, first={Laser Interferometer Gravitational Wave Observatory \citep[LIGO;][]{1992Sci...256..325A}}}
\newglossaryentry{aligo}{name=Advanced LIGO, description={Advanced LIGO}, sort=ligo2}
\newglossaryentry{lisa}{name=LISA, description={Laser Interferometer Space Antenna \citep{1994ESAJ...18..219J}}, first={Laser Interferometer Space Antenna \citep[LISA;][]{1994ESAJ...18..219J}}}
\newglossaryentry{ccd}{name=CCD,description={Charged Coupled Device}, first={charged coupled device (CCD)}, firstplural={charged coupled devices (CCDs)}}

\newcommand{\sn}[2]{SN~#1#2}

\newcommand{\snr}[1]{SNR~#1}

\newglossaryentry{sn}{name=Supernova, text={SN}, plural={SNe}, description={exploding star}, nonumberlist=true}
\newglossaryentry{snia}{name=Type~Ia (SN~Ia), text={SN~Ia}, description={Thermonuclear explosion of a white dwarf - spectra show no hydrogen but a strong silicon line},first={Type~Ia supernova (SN~Ia)}, firstplural={Type Ia supernovae (SNe~Ia)}, plural={SNe~Ia}, parent=sn, nonumberlist=true}
\newcommand{\sneia}{\glspl*{snia}}
\newcommand{\snia}{\gls*{snia}}

\newglossaryentry{branchnormal}{name={branch-normal}, text=\textit{Branch-normal}, description={Large homogeneous class of Type Ia Supernovae, defined in \citet{1993AJ....106.2383B}}, first={\textit{Branch-normal} SNe Ia \citep{1993AJ....106.2383B}}, parent=snia} 
\newglossaryentry{91t}{name={91T-like}, description={Luminous class of Type Ia supernovae similar to \sn{1991}{T} \citep{1992AJ....103.1632P}} , first={91T-like}, parent=snia} 
\newglossaryentry{91bg}{name={91bg-like}, description={Faint class of Type Ia supernovae similar to \sn{1991}{bg} \citep{1992AJ....104.1543F}}, first={91bg-like}, parent=snia} 
\newglossaryentry{02cx}{name={02cx-like}, description={Peculiar class of Type Ia supernovae similar to \sn{2002}{cx} \citep{2003PASP..115..453L}}, first={02cx-like \sneia\ \citep{2003PASP..115..453L}}, parent=snia} 

\newglossaryentry{snibc}{name=Type~Ib/c, text={SN~Ib/c}, description={Collapse of the core of a massive star -  spectrum shows no hydrogen and no silicon line},first={Type~Ib/c supernova (SN~Ib/c)}, firstplural={Type~Ib/c supernovae (SNe~Ib/c)}, plural={SNe~Ib/c}, parent=sn}

\newglossaryentry{snib}{name=Type~Ib, text={SN~Ib}, description={Spectrum shows no hydrogen and no silicon, but helium line},first={Type Ib supernova (SN~Ib)}, firstplural={Type~Ib supernovae (SNe~Ib)}, plural={SNe~Ib}, parent=snibc}

\newglossaryentry{snic}{name=Type~Ic, text={SN~Ic}, description={Spectrum shows no hydrogen, no silicon and no helium line},first={Type~Ic supernova (SN~Ic)}, firstplural={Type~Ic supernovae (SNe~Ic)}, plural={SNe~Ic}, parent=snibc}

\newglossaryentry{snii}{name=Type~II, text={SN~II}, description={Collapse of the core of a massive star - spectrum shows strong hydrogen line},first={Type~II supernova (SN~II)}, firstplural={Type~II supernovae (SNe~II)}, plural={SNe~II}, parent=sn}

\newglossaryentry{sniib}{name=Type~IIb, text={SN~IIb}, description={Spectrum shows hydrogen and helium lines},first={Type~IIb supernova (SN~IIb)}, firstplural={Type~IIb supernovae (SNe~IIb)}, plural={SNe~IIb}, see=snib, parent=snii}

\newglossaryentry{sniip}{name=Type~II~Plateau (Type IIP), text={SN~IIP}, description={Lightcurve shows plateau},first={Type~IIP supernova (SN~IIP)}, firstplural={Type~II Plateau supernovae \citep[SNe~IIP;][]{1979A&A....72..287B}}, plural={SNe~IIP}, parent=snii}

\newglossaryentry{sniil}{name=SN~II~Linear, text={SN~IIL}, description={Lightcurve shows no plateau, but linear decline},first={Type~IIL supernova (SN~IIL)}, firstplural={Type~II~Linear supernovae \citep[SNe~IIL;][]{1990MNRAS.244..269S}}, plural={SNe~IIL}, parent=snii}

\newglossaryentry{sniin}{name=Type II narrow-lined (Type IIn), description={Spectrum shows narrow lines},first={Type~II~narrow-lined supernova (SN IIn)}, firstplural={Type~IIn supernovae (SNe~IIn)}, plural={SNe~IIn}, parent=snii}

\newglossaryentry{snr}{name=Remnant (SNR), text=SNR, description={Remnant left visible post-explosion}, first={supernova remnant (SNR)}, firstplural={supernova remnants (SNRs)}, parent=sn}

\newglossaryentry{dtd}{name=DTD,description={delay time distribution - expected supernova rate over time after a brief outburst of starformation},first={delay time distribution (DTD)}, firstplural={delay time distributions (DTDs)}, plural=DTDs}

\newglossaryentry{hvg}{name=HVG,description={high velocity gradient - Type Ia supernovae with a fast evolution of photospheric velocity},first={high velocity group (HVG)}, firstplural={high velocity groups (HVGs)}, plural=HVGs, parent=snia}

\newglossaryentry{lvg}{name=LVG,description={low velocity gradient - Type Ia supernovae with a slow evolution of photospheric velocity},first={low velocity group (LVG)}, firstplural={low velocity groups (LVGs)}, plural=LVGs, parent=snia}

\newglossaryentry{wd}{name=white dwarf (WD), text=WD, description={White Dwarf - extremely dense stellar remnant}, first={white dwarf (WD)}}
\newglossaryentry{onemgwd}{name= Oxygen/Neon (ONe), text={ONe-WD},description={Oxygen/Neon White Dwarf}, first={oxygen/neon White Dwarf (ONe-WD)}, parent=wd}
\newglossaryentry{cowd}{name=carbon/oxygen (CO), text={CO-WD}, description={carbon/oxygen white dwarf}, first={carbon/oxygen white dwarf (CO-WD)}, firstplural = {carbon/oxygen white dwarfs (CO-WDs)}, parent=wd}

\newglossaryentry{sds}{name=SD-Scenario,description={single-degenerate scenario (single white dwarf accreting from non-degenerate companion)}, first={single-degenerate scenario (SD-scenario)}}

\newglossaryentry{dds}{name=DD-Scenario, description={double degenerate scenario (merging of two white dwarfs)}, first={double-degenerate scenario (DD-scenario)}}

\newglossaryentry{sss}{name=SSS, text={supersoft \xray\ source}, description={supersoft \xray\ source - believed to be emitted by nuclear fusion on a white dwarf's surface}}

\newglossaryentry{amcvn}{name=AM CVn, description={AM Canum Venaticorum star (white dwarf accreting hydrogen poor matter from a companion star; see \cite{2005ASPC..330...27N})}}

\newglossaryentry{rlof}{name=RLOF, description={Roche Lobe Overflow (see \citet{1971ARA&A...9..183P} for a more detailed description)}, first={Roche-lobe overflow (RLOF)}}

\newglossaryentry{mchan}{name={Chandrasekhar mass~}, text={Chandrasekhar~mass}, symbol={\ensuremath{M_\textrm{Chan}}}, plural={Chandrasekhar~masses}, description={Mass when the core of a star collapses due to insufficient degeneracy pressure - for a white dwarf $\approx1.38\,M_\odot$ see \citet{1931ApJ....74...81C}}, first={Chandrasekhar~mass \citep[$M_\textrm{Chan}=1.38\,M_\odot$;][]{1931ApJ....74...81C}}, sort=mchan}

\newglossaryentry{w7}{name={W7 model},description={W7 model \citep{1984ApJ...286..644N}},first = {W7 model \citep{1984ApJ...286..644N}}}

\newcommand{\stara}{\object[{[RCM2004] Tycho A}]{\hbox{Tycho-A}}}

\newcommand{\starb}{\object[{[RCM2004] Tycho B}]{\hbox{Tycho-B}}}
\newcommand{\starc}{\object[{[RCM2004] Tycho C}]{\hbox{Tycho-C}}}
\newcommand{\stard}{\object[{[RCM2004] Tycho D}]{\hbox{Tycho-D}}}
\newcommand{\stare}{\object[{[RCM2004] Tycho E}]{\hbox{Tycho-E}}}
\newcommand{\starg}{\object[{[RCM2004] Tycho G}]{\hbox{Tycho-G}}}

\newglossaryentry{rp04}{name=RP04, text={RP04}, description={short for \citet{2004Natur.431.1069R}}, first={\citet[][henceforth RP04]{2004Natur.431.1069R}}}
\newcommand{\rl}{\gls{rp04}}
\newglossaryentry{gh09}{name=GH09, text={GH09}, description={short for \citet{2009ApJ...691....1G}}, first={\citet[][henceforth GH09]{2009ApJ...691....1G}}}
\newcommand{\gh}{\gls{gh09}}
\newglossaryentry{wek09}{name=WEK09, text={WEK09}, description={short for \citet{2009ApJ...701.1665K}}, first={\citet[][henceforth WEK09]{2009ApJ...701.1665K} }}
\newcommand{\wek}{\gls{wek09}}

\newglossaryentry{ew}{name=Equivalent Width, text={EW}, description={width of a rectangle that has the same area as a spectral line when taken to zero flux}, first={equivalent width (EW)}, firstplural={equivalent widths (EWs)}}
\newglossaryentry{agb}{name=AGB,description={Asymptotic Giant Branch}}
\newglossaryentry{cmb}{name=CMB,description={Cosmic Microwave Background}}
\newglossaryentry{csm}{name=CSM,description={Circumstellar Medium}, first={circumstellar medium (CSM)}}
\newglossaryentry{csi}{name=CSI,description={Circumstellar Interaction}, first={circumstellar interaction (CSI)}}
\newglossaryentry{ism}{name=ISM,description={Interstellar Medium}, first={interstellar medium (ISM)}}
\newglossaryentry{ige}{name=IGE,description={Iron Group Element}, first={iron group element (IGE)}, firstplural={iron group elements (IGEs)}}
\newglossaryentry{epm}{name=EPM,description={Expanding Photosphere Method \citep{1974ApJ...193...27K}}, first={Expanding Photosphere Method (EPM)}}
\newglossaryentry{aic}{name=AIC,description={Accretion Induced Collapse}, first={accretion induced collapse (AIC)}}
\newglossaryentry{ime}{name=IME,description={Intermediate Mass Element}, first={intermediate mass element (IME)}, firstplural={intermediate mass elements (IMEs)}}
\newglossaryentry{h0}{name=\ensuremath{H_0},description={Hubbles constant}}
\newglossaryentry{nse}{name=NSE,description={Nuclear Statistical Equilibrium}, first={nuclear statistical equilibrium (NSE)}}
\newglossaryentry{cdm}{name=CDM,description={Cold Dark Matter}}
\newglossaryentry{grb}{name=GRB,description={Gamma Ray Burst}, first={Gamma Ray Burst (GRB)}, firstplural={Gamma Ray Bursts (GRBs)}}
\newglossaryentry{donor}{name=donor,description={non-degenerate companion in the \gls{sds}}}
\newglossaryentry{mainsequence}{name=main sequence,description={main sequence star}}
\newglossaryentry{redgiant}{name=red giant,description={red giant star}}
\newglossaryentry{mlcs}{name=MLCS,description={Multicolor Light Curve Shape method \citep[MLCS;][]{1996ApJ...473...88R}}, first={Multicolor Light-Curve Shape method \citep[MLCS;][]{1996ApJ...473...88R}}}
\newglossaryentry{rsoph}{name=RS~Ophiuci ,description={white dwarf accreting from a red giant - assumed progenitor of the \gls{sds}}, sort=rsoph}
\newglossaryentry{usco}{name=U~Scorpii,description={white dwarf accreting from a main sequence star - assumed progenitor of the \gls{sds}}, sort=usco}
\newglossaryentry{rcw86}{name=RCW~86,description={supernova remnant sometimes associated with \sn{185}{}}, sort=rcw86}
\newglossaryentry{casa}{name=Cas~A,description={Cassiopeia A supernova remnant - probably a \gls{snib} event}}
\newglossaryentry{cepheid}{name=Cepheid,description={very luminous variable star with a strong luminosity period relationship}}
\newglossaryentry{urca}{name=Urca, text=\textit{Urca}, description={process predominatly contributing to cooling in stars. The \textit{Urca} process consists of alternating electron-capture and $\beta^{-}$ decay of two nuclei pairs.},sort=urca} 
\newglossaryentry{alphacen}{name=Alpha Centauri,description={one of the brightest stars in the night sky and a close binary}}
\newglossaryentry{pcygni}{name={P Cygni}, text={P Cygni},description={a hypergiant luminous blue variable with strong winds. Often referred to as a description for their line profiles showing a emission peak at the rest wavelength of the line and a blue-shifted absorption trough.}}

\newglossaryentry{teff}{name={effective temperature~}, text={effective temperature}, symbol={\ensuremath{T_\textrm{eff}}}, description={Temperature of a blackbody emitting the same total energy}, sort=teff}

\newglossaryentry{logg}{name={surface gravity~}, text={surface gravity}, symbol={\ensuremath{\textrm{log}\,g}}, description={gravity at the surface of a star}, sort=logg}
\newglossaryentry{feh}{name={metallicity~}, text={metallicity}, symbol=\textrm{[Fe/H]},description={iron abundance relative to the sun}, sort=feh}

\newglossaryentry{texp}{name={time since explosion~}, text={time since explosion}, text={time since explosion}, symbol={\ensuremath{t_{\rm exp}}},description={time since explosion (measured in days)}, sort=texp, first={time since explosion (\ensuremath{t_{\rm exp}})}}

\newglossaryentry{lmc}{name=LMC,description={Large Magellanic Cloud}, first={Large Magellanic Cloud (LMC)}, sort=lmc}
\newglossaryentry{smc}{name=SMC,description={Small Magellanic Cloud}, sort=smc}
\newglossaryentry{z}{name=\ensuremath{z},description={redshift}, sort=z}
\newcommand{\teff}{\glssymbol*{teff}}
\newcommand{\logg}{\glssymbol*{logg}}
\newcommand{\feh}{\glssymbol*{feh}}

\renewcommand{\sn}[2]{\object{SN~#1#2}}
\usepackage{amsmath}
\usepackage{natbib}

\newcommand{\fei}{\ion{Fe}{1}}
\newcommand{\feii}{\ion{Fe}{2}}
\shortauthors{W.E. Kerzendorf et al.}
\shorttitle{Progenitor search in Tycho's SN}

\newcommand{\plotdir}{./}

\begin{document}
\title{A High-Resolution Spectroscopic Search for the Remaining Donor for Tycho's Supernova}
\author{Wolfgang~E.~Kerzendorf\altaffilmark{1,2}, David~Yong\altaffilmark{1}, Brian~P.~Schmidt\altaffilmark{1},  Joshua~D.~Simon\altaffilmark{7}, C.~Simon~Jeffery\altaffilmark{3}, Jay~Anderson\altaffilmark{4}, Philipp~Podsiadlowski\altaffilmark{5}, Avishay~Gal-Yam\altaffilmark{6},
Jeffrey~M.~Silverman\altaffilmark{8}, Alexei~V.~Filippenko\altaffilmark{8}, Ken'ichi~Nomoto\altaffilmark{9,10}, Simon~J.~Murphy\altaffilmark{1}, Michael~S.~Bessell\altaffilmark{1},
		Kim~A.~Venn\altaffilmark{11}, and Ryan~J.~Foley\altaffilmark{12}} 
\email{wkerzend@mso.anu.edu.au}

\altaffiltext{1}{Research School of Astronomy and
Astrophysics, Mount Stromlo Observatory, Cotter Road, Weston Creek,
ACT 2611, Australia}
\altaffiltext{2}{Department of Astronomy and Astrophysics, University of Toronto, 50 Saint George Street, Toronto, ON M5S 3H4, Canada}
\altaffiltext{3}{Armagh Observatory, College Hill, Armagh BT61 9DG, Northern Ireland}

\altaffiltext{4}{Space Telescope Science Institute, Baltimore, MD 21218, USA}
 
\altaffiltext{5}{Department of
Astrophysics, University of Oxford, Oxford, OX1 3RH, United
Kingdom} 

\altaffiltext{6}{Benoziyo Center for Astrophysics, Faculty of Physics, The Weizmann Institute of Science, Rehovot 76100, Israel} 

\altaffiltext{7}{Observatories of the Carnegie Institution of Washington, 813 Santa Barbara St., Pasadena, CA 91101, USA}

\altaffiltext{8}{Department of Astronomy, University of California, Berkeley, CA 94720-3411, USA}

\altaffiltext{9}{Kavli Institute for the Physics and Mathematics of the Universe,
 The University of Tokyo, 5-1-5 Kashiwanoha, Kashiwa, Chiba 277-8583,
 Japan}

\altaffiltext{11}{Department of Physics and Astronomy, University of Victoria, Elliott Building, 3800 Finnerty Road, Victoria, BC V8P 5C2, Canada}

\altaffiltext{12}{Harvard-Smithsonian Center for Astrophysics, 60 Garden Street, Cambridge, MA 02138, USA}

\begin{abstract}
In this paper, we report on our analysis using {\it Hubble Space Telescope} astrometry and Keck-I HIRES spectroscopy of the central six stars of Tycho's supernova remnant (SN~1572). With these data, we measured  the proper motions, radial velocities, rotational velocities, and chemical abundances of these objects. Regarding the chemical abundances, we do not confirm the unusually high [Ni/Fe] ratio previously reported for Tycho-G. Rather, we find that for all metrics in all stars, none exhibit the characteristics expected from traditional SN~Ia single-degenerate-scenario calculations. The only possible exception is Tycho-B, a rare, metal-poor A-type star; however, we are unable to find a suitable scenario for it. Thus, we suggest that SN~1572 cannot be explained by the standard single-degenerate model. 
\end{abstract}


\maketitle

\section{Introduction}
\label{sec:sn1572_hires:introduction}

\sneia\ are of great interest. They represent some of the most extreme physical situations in stellar astronomy, control the chemical evolution of galaxies and the Universe at intermediate to late times by producing large amounts of iron-group elements, and are uniquely powerful cosmic distance probes. But despite their wide-ranging significance, fundamental uncertainties remain around the progenitors of these cataclysmic events. 

There is general consensus that \sneia\ are caused by the deflagration/detonation of a \gls{cowd} which is accreting material from a binary companion. Scenarios exist where the explosion can be initiated from a detonation on the surface of the star \citep{1995ApJ...452...62L, 2010A&A...514A..53F}, through runaway carbon burning in the white dwarf's interior, or through a cataclysmic merger of objects.

Observationally, two main models for this accretion process can be identified. The \gls{sds} sees the accretion process occurring through \gls{rlof} of a close nondegenerate companion (also known as a \gls{donor} star). This companion, which has undergone common-envelope evolution with the white dwarf, can be a helium, main-sequence, subgiant, or red giant star. In all cases the donor star should survive the explosion (except for possibly in the case of the helium-star donor; R.~Pakmor 2012, private communication) and remain visible post-explosion.

The second scenario is the  dynamical merger of two white dwarfs, known as the \gls{dds}. In this scenario, the coevolution of two stars eventually leads to a close binary of two white dwarfs, which are able, through the emission of gravitational radiation, to merge over a wide range of times after the initial formation of the system. In most cases this would leave no remaining star \citep[e.g.,][]{2010Natur.463...61P}.

Both scenarios have support in observations and theory. The detection of circumstellar material around certain \sneia\ \citep{2007Sci...317..924P,2009ApJ...702.1157S, 2011Sci...333..856S, 2012arXiv1203.2916F} provides evidence for the \gls{sds}. On the other hand, the lack of substantial hydrogen in the majority of other \sneia\ \citep{2007ApJ...670.1275L} poses a challenge to the \gls{sds}.

\citet{2010ApJ...708.1025K} suggests that the interaction of the shock wave with the nondegenerate companion should result in a light excess at early times of an \snia~light curve, which depends on the viewing angle and the companion radius. Such an excess has not yet been observed \citep{2010ApJ...722.1691H, 2011Ap&SS.tmp...40T, 2011arXiv1106.4008B,2011MNRAS.416.2607G}, which is at odds with \gls{redgiant} companions forming the majority of \sneia. \citet{2011ApJ...730L..34J}, \citet{2011ApJ...738L...1D}, and \citet{2012ApJ...756L...4H,2012ApJ...744...69H}, however, have suggested a scenario where the white dwarf is spinning and thus can accrete above the Chandrasekhar limit. The explosion would only occur once the white dwarf had spun down sufficiently, which would give the red giant a chance to evolve and would not require the detection of the early excess in the light curve in a red giant progenitor scenario.

Population-synthesis calculations are challenging, with various authors getting different results for the same inputs. However, there is a general trend from these calculations that neither single-degenerate nor double-degenerate stars can provide enough systems to explain the observed \snia\ rate \citep{2008ApJ...677L.109H,2009ApJ...699.2026R, 2010A&A...515A..89M,2010A&A...521A..85Y}. Several authors suggest that the population might comprise both single-degenerate and double-degenerate systems.

The physics of white-dwarf mergers is challenging to simulate numerically, but in the simplest calculations, these mergers lead to the formation of a neutron star via electron capture, rather than to a thermonuclear explosion \citep{1985A&A...150L..21S}. Recently, \citet{2010Natur.463...61P} have shown that for certain parameters (white-dwarf binaries with a mass ratio very close to 1) the merger may explain subluminous supernovae \citep[e.g., \sn{1991}{bg}; see][for a review]{1997ARA&A..35..309F}, although \citet{2011arXiv1101.5132D} note that the initial conditions of the system may change these conclusions.

SN~2011fe was detected only $\sim 11$\,hr after the explosion, and (with a distance of 6.4\,Mpc) is one of the closest \sneia\ ever found \citep{2011Natur.480..344N}. \citet{2011Natur.480..344N} and \citet{2011arXiv1110.2538B} have not found any early-time light-curve excess predicted by \citet{2010ApJ...708.1025K} and thus rule out a red giant donor. Radio and X-ray observations by \citet{2011arXiv1109.2912H} show no strong signs of pre-explosion outflows, which again contradicts a red giant scenario for SN~2011fe. Additional radio measurements by \cite{2012arXiv1201.0994C} suggest a low density around SN~2011fe, which is at odds with many conventional single-degenerate scenarios. \citet{2011Natur.480..348L} have searched pre-explosion archival images and can also rule out luminous red giants and almost all helium stars as donors. In addition, \citet{2012ApJ...744L..17B} have used images believed to have been taken 4\,hr post-explosion and suggest the companion radius to be less than 0.1\,R$_\odot$. Most of these results are consistent with a main-sequence companion or white-dwarf companion.

Because it is very difficult to obtain robust constraints on the progenitor system in the immediate aftermath of a $10^{51}$\,erg explosion, an alternative is to study somewhat older and more nearby SNe that can be observed in great detail. \rl\ have tried to directly detect donor stars in old and nearby \snia\ remnants within the Milky Way. They have identified two historical Galactic \glspl{sn} well suited to this task --- \sn{1006}{}\ and \sn{1572}{} (Tycho's SN). Both remnants are young (1000 and 440\,yr old, respectively), closeby ($2.2\pm0.08\,\textrm{\kpc}$, \citealt{2003ApJ...585..324W}; $2.8\pm0.8\,\textrm{\kpc}$, \citealt{2004ApJ...612..357R}, respectively), almost certainly \sneia\ from their observational signatures \citep{2006ApJ...645.1373B,2004ApJ...612..357R, 2008Natur.456..617K, 2008ApJ...681L..81R}, and not overwhelmed by Galactic extinction. In this paper, we will focus on \sn{1572}{}. 

\rl\ investigated most bright stars in the central region of \sn{1572}{}\ and found a star with an unusual spatial motion (\starg\ by their nomenclature); they suggested this as a possible donor star for \sn{1572}{}. While the star has an unusual spatial motion compared to other stars in the field, its current location and proper motion place it a significant distance from the center of the supernova remnant (SNR) --- a feature difficult to explain in connecting \starg\ to \snr{1572}{}. 

One consequence of \gls{rlof} is a rotational velocity induced on the \gls{donor} star by tidal locking in the system. This results in an unusually large rotational velocity, related to the orbital velocity of the binary system, and it can be used to single out possible donor stars from nearby unrelated stars. \wek\ investigated the rotation of \starg\ but found no excess rotational velocity compared to a normal star. A comparison of \wek's measurements of \starg, including a revised radial velocity $\vrad$, with Galactic kinematic models showed that it is statistically consistent with an unrelated thick/thin-disk star. However, \wek\ were able to provide an {\it a priori} unlikely donor-star scenario, where the star was able to lose its rotational signature. 


\gh\ analyzed a spectrum of \starg\ observed with the \gls{hires} instrument on the Keck-I 10\,m telescope, finding a $\vrad$ value consistent with \wek's revised $\vrad$. They also measured the stellar parameters and metallicity of \starg, concluding that it has an unusually high nickel abundance, which they claim can be attributed to the accretion of ejecta material on the \gls{donor} star during the explosion. 

In this paper we analyze \gls{hires} spectra of the six bright stars near the center of \snr{1572}{}. These spectra were taken as part of the same program that obtained the data used by \gh, and we independently reanalyze the \starg\ spectrum in our program. We describe the observational data and our data-reduction procedures in \S \ref{sec:sn1572_hires:observ-data-reduct}. Section \ref{sec:analysis} is divided into six subsections detailing the measurements of proper motion, radial velocity, rotation, stellar parameters, and abundances, and we provide a detailed comparison between our and \gh's measurements of \starg. In \S \ref{sec:sn1572_hires:discussion} we analyze the measurements of each star to investigate its potential association with \snr{1572}, and we present our conclusions in \S \ref{sec:sn1572_hires:conclusion}.

\section{Observations and Data Reduction}
\label{sec:sn1572_hires:observ-data-reduct}

We obtained spectra with the \gls{hires} spectrograph on the Keck-I telescope on Mauna Kea. The observations were made on 2006 September 10 and 2006 October 11 UT. Slits B5 and C1 (width 0.86\arcsec; B5 length 3.5\arcsec, C1 length 7.0\arcsec) were used, resulting in wavelength coverage of 3930--5330\,\AA, 5380--6920\,\AA, and 6980--8560\,\AA\ with $R = \lambda/\Delta\lambda \approx$ 50,000, providing us with the necessary spectral resolution and wavelength coverage to determine stellar parameters.
 
The spectra were reduced using the \gls{makee} package. All spectra were corrected to heliocentric velocities, using the \gls{makee} sky-line method. The spectra were not corrected for telluric absorption lines, but only regions known to be free from telluric contamination were used in the analysis to derive the stellar parameters. The final exposure times of the combined spectra for each candidate and the signal-to-noise ratio (S/N) at 5800--5900\,\AA\ are shown in Table~\ref{tab:candexpo}. Finally, we normalized the spectrum using the \gls{iraf} task \textsc{continuum}. We note that \starc\ and \stard\ were observed on the same slit (C1); they are separated by 2.1\arcsec, and the seeing was $\sim 0.8\arcsec$, with \starc\ being roughly five times brighter than \stard.

\begin{deluxetable}{lccccccc}
\tablecolumns{7}
\tablecaption{Observations \label{tab:candexpo}}
\tablehead{%
\colhead{Tycho} &      
\colhead{$\alpha$(J2000)} &   		      
\colhead{$\delta$(J2000)} &               
\colhead{Date} &                 
\colhead{Slit} &                    
\colhead{$t_\textrm{exp}$} &
\colhead{$V$\tablenotemark{a}}&                    
\colhead{S/N\tablenotemark{b}} \\                   
\colhead{(Name)}&%
\colhead{(hh:mm:ss.ss)}&%
\colhead{(dd:mm:ss.ss)}&%
\colhead{(dd/mm/yy)}&%
\colhead{}&%
\colhead{(s)}&%
\colhead{(mag)}&%
\colhead{}%
\\
}

\startdata 
A & 00:25:19.73 & +64:08:19.60 & 10/09/06 & B5& 900 &13,29&$ \sim 48$ \\
B & 00:25:19.95 & +64:08:17.11 & 10/09/06 & B5 & 1200 &15.41&$ \sim 45$ \\ 
C & 00:25:20.40 & +64:08:12.32 & 11/10/06 & C1 &  10,800&19.06\tablenotemark{c} & $ \sim 8$\\ 
D & 00:25:20.60 & +64:08:10.82 & 11/10/06 & C1 & 10,800 &20.70& $ \sim 3$ \\
E & 00:25:18.29 & +64:08:16.12 & 11/10/06 & C1 & 9000 &19.79& $\sim 9$\\ 
G & 00:25:23.58 & +64:08:02.06 & 10/09/06 \& 11/10/06  & B5\&C1 & 24,000 & 18.71& $\sim 25$ \\
\enddata
\tablenotetext{a}{Magnitudes from \rl.}
\tablenotetext{b}{The S/N value was obtained by measuring the root-mean square of the pixels (each resolution element is sampled by 2 pixels) in continuum regions near 5800--5900\AA. For the purposes of measuring the stellar parameters, the spectrum was convolved and the S/N increases by a factor of 2.24.}

\tablenotetext{c}{\rl\ notes that this is an unresolved pair, with a brighter bluer component ($V=19.38$ mag) and a fainter redder component ($V=20.53$ mag).}
\end{deluxetable}

In addition, we obtained low-resolution spectra ($R \approx 1200$) of
\starb\ with the dual-arm \gls{lris} mounted on the Keck-I telescope. The
data were taken on 2010 November 7 UT, using only the blue
arm with the 600/4000 grism and the 1\arcsec-wide slit. This resulted
in a wavelength coverage of 3200--5600\,\AA. These data
were taken to obtain a precise  measurement of the surface gravity for \starb\ using the size of the Balmer decrement \citep{2007PASP..119..605B}.

The spectrum of \starb\ was reduced using standard techniques \citep[e.g.,][]{Foley03}. Routine CCD processing and spectrum extraction
were completed with \gls{iraf}, and the data were extracted with
the optimal algorithm of \citet{Horne86}. We obtained the wavelength
scale from low-order polynomial fits to the calibration-lamp spectra.
Small wavelength shifts were then applied to the data after measuring the offset by
cross-correlating a template sky to the night-sky lines that were
extracted with the star. Using our own \gls{idl} routines, we fit a
spectrophotometric standard-star spectrum to the data in order to flux
calibrate \starb\ and remove telluric lines \citep{Horne86,Matheson00}.

\section{Analysis}
\label{sec:analysis}

\subsection{Astrometry}
\label{sec:propmot}
Proper motions can be used to identify potential donor stars because donor stars freely travel with their orbital velocity after the SN explosion disrupts the system. \rl\ suggested \starg\ as a possible donor due to its unusually high values for both the proper motion and radial velocity. For this work we measured proper motions for 201 stars within one arcminute of the remnant's center. We used archival {\it Hubble Space Telescope (HST)} images for three different epochs ({\it HST} Programs GO-9729 and GO-10098; November 2003, August 2004, May 2005), each consisting of three exposures (1\,s, 30\,s, and 1440\,s) with the F555W filter using the Advanced Camera for Surveys (ACS). The scale in each exposure is 50\,mas\,pixel$^{-1}$. This dataset results in a maximum baseline of 18 months. 

We used an image from the middle epoch (2004) to establish a reference frame and oriented the pixel coordinate system with the equatorial system. We then applied a distortion correction for the F555W filter \citep{2006acs..rept....1A} and calculated transformations between all other images and the reference image. Next we used these transformations to calculate the position of all stars in the reference coordinate system, with the overall uncertainty of each position estimated.
Some faint stars were not detected in the shorter exposures and were thus excluded from proper-motion measurements. In total, 114 stars were used in the astrometric analysis.

For each star, we fit a linear regression for the stellar positions over time in the pixel coordinates (which were aligned with the equatorial system). The $x$ and $y$ data were treated as independent measurements, with separate regressions solved for each axis direction. Uncertainties were estimated using standard least-squares analysis and the individual uncertainty estimates of each object's positions.

There are three J2000 measurements of the geometric center of \sn{1572}{}\ from different datasets. \cite{1997ApJ...491..816R} used \gls{vla} data to measure the center as \rasc{00}{25}{14}{95}, \decl{+64}{08}{05.7};  \citet{2000ApJ...545L..53H} used \gls{rosat} data to measure \rasc{00}{25}{19}{0}, \decl{+64}{08}{10}{}; and \cite{2005ApJ...634..376W} used \gls{chandra} data to measure \rasc{00}{25}{19}{40}, \decl{64}{08}{13.98}. We note that the X-ray centers agree rather well with a difference of less than 5\arcsec, but the radio center is located roughly 30\arcsec away from the X-ray centers. Thus, we believe the error in the geometric center to be rather large (of order 30\arcsec).

Table~\ref{tab:propmot} lists the proper motions and uncertainties of all stars mentioned by \rl\ (19 stars) which were analyzed in this work, as well as the distance to the geometric \xray\ center measured by \gls{chandra}.

We note that Tycho-2 has a relatively high proper motion, but its position in the year 1572 was 67.95\arcsec\ away from the remnant's center, and we thus exclude it as a viable candidate for the donor.

\begin{figure*}[ht!] 
   \centering
   \includegraphics[width=\textwidth]{\plotdir 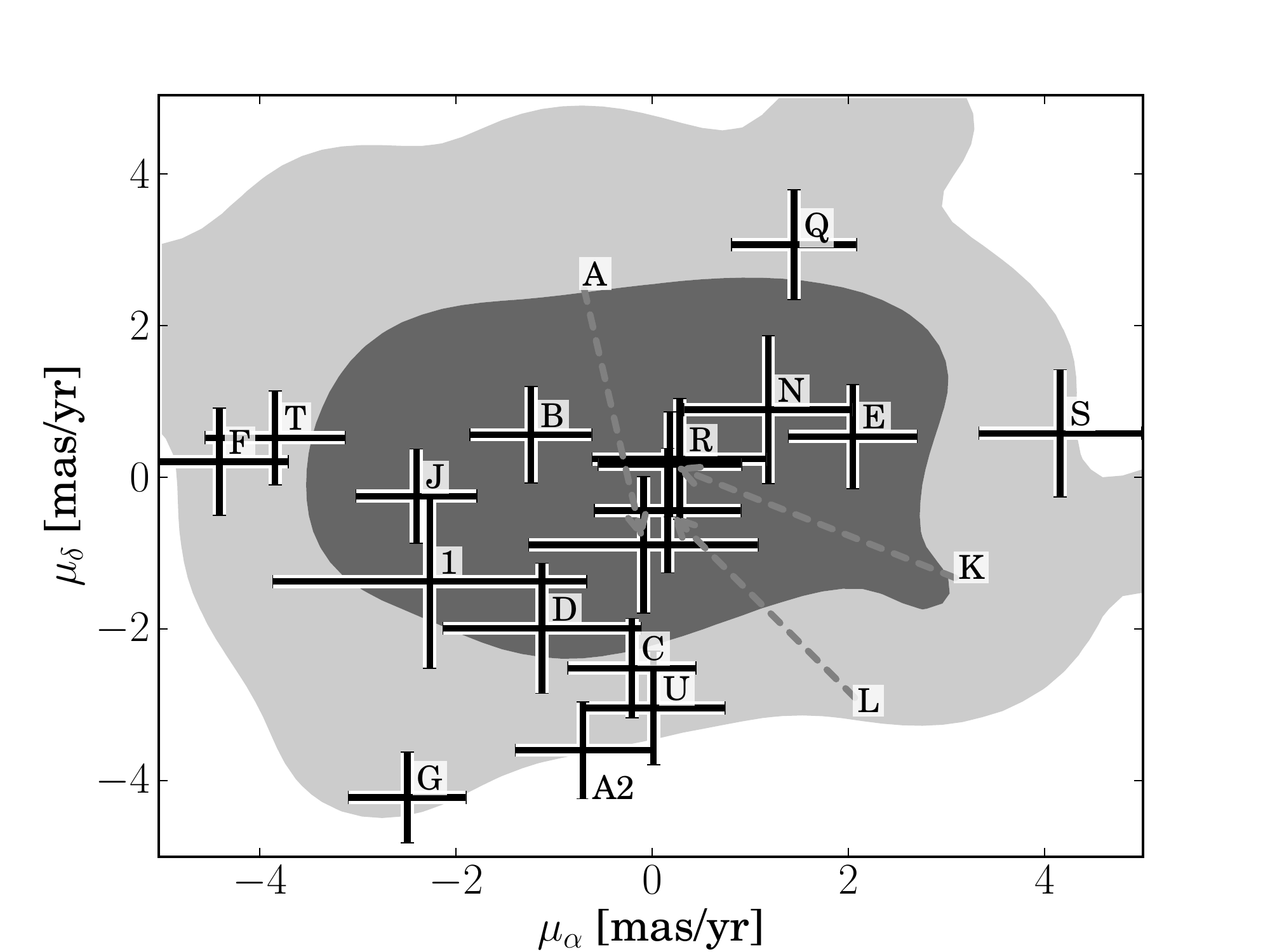}
   \caption[Proper-motion measurement of stars in SNR~1572 using only {\it HST} images]{The contours show the distribution of proper motions (68\% and 95\% probability) for all stars measured toward the Tycho SNR --- excluding the named stars.
    We show the location of the candidate stars and their uncertainties on top of this distribution. Tycho-2  (called HP-1 in \wek) is not shown in this figure as it is an extreme outlier with $\mu_\alpha=75$\,\masyr\ and $\mu_\delta=-4.4$\,\masyr; it is also at a large distance from the remnant's geometric center (46\arcsec). We find, using the Besan\c{c}on model as a proxy, that the contamination of this sample with foreground objects (less than 2\,kpc) is less than 15\%.}
   \label{fig:propmot_sn1572_hires}
\end{figure*}

In Figure \ref{fig:propmot_sn1572_hires}, we compare the distribution of proper motions of all measured stars to our candidates. All of our candidates are reconcilable with a normal proper-motion distribution.

\begin{deluxetable}{lccccccccc}
\tablecaption{Proper Motions of Candidates\label{tab:propmot}}
\tablecolumns{10} 
\tablehead{ 
\colhead{Tycho} &          
\colhead{$\alpha$(J2000)} &   		      
\colhead{$\delta$(J2000)} &
\colhead{$\mu_\alpha$} &
\colhead{$\mu_\delta$} & 
\colhead{$\Delta\mu_\alpha$} & 
\colhead{$\Delta\mu_\delta$} & 
\colhead{$\mu_l$}& 
\colhead{$\mu_b$}& 
\colhead{$r$}\\%
\colhead{(Name)}&%
\colhead{(hh:mm:ss.ss)}&%
\colhead{(dd:mm:ss.ss)}&%
\multicolumn{2}{c}{($\masyr$)}&
\multicolumn{2}{c}{($\masyr$)}&
\multicolumn{2}{c}{($\masyr$)}&
\colhead{(\arcsec)}\\
}
\startdata
B & 0:25:19.97 & 64:08:17.1 & -1.24 & 0.56 & 0.62 & 0.64 & -1.17 & 0.68 & 4.86\\
A & 0:25:19.73 & 64:08:19.8 & -0.09 & -0.89 & 1.17 & 0.90 & -0.18 & -0.88 & 6.21\\
A2 & 0:25:19.81 & 64:08:20.0 & -0.71 & -3.60 & 0.69 & 0.64 & -1.07 & -3.51 & 6.58\\
C & 0:25:20.38 & 64:08:12.2 & -0.21 & -2.52 & 0.65 & 0.65 & -0.46 & -2.48 & 6.66\\
E & 0:25:18.28 & 64:08:16.1 & 2.04 & 0.54 & 0.66 & 0.69 & 2.09 & 0.33 & 7.60\\
D & 0:25:20.62 & 64:08:10.8 & -1.12 & -1.99 & 1.01 & 0.86 & -1.32 & -1.87 & 8.60\\
1 & 0:25:16.66 & 64:08:12.5 & -2.27 & -1.37 & 1.60 & 1.15 & -2.40 & -1.14 & 18.00\\
F & 0:25:17.09 & 64:08:30.9 & -4.41 & 0.20 & 0.70 & 0.71 & -4.37 & 0.65 & 22.69\\
J & 0:25:15.08 & 64:08:05.9 & -2.40 & -0.25 & 0.62 & 0.62 & -2.42 & -0.00 & 29.44\\
G & 0:25:23.58 & 64:08:01.9 & -2.50 & -4.22 & 0.60 & 0.60 & -2.91 & -3.95 & 29.87\\
R & 0:25:15.51 & 64:08:35.4 & 0.28 & 0.24 & 0.89 & 0.80 & 0.30 & 0.21 & 33.23\\
N & 0:25:14.73 & 64:08:28.1 & 1.18 & 0.89 & 0.86 & 0.98 & 1.27 & 0.77 & 33.66\\
U & 0:25:19.24 & 64:07:37.9 & 0.01 & -3.04 & 0.73 & 0.75 & -0.30 & -3.03 & 36.06\\
Q & 0:25:14.81 & 64:08:34.2 & 1.45 & 3.07 & 0.64 & 0.72 & 1.75 & 2.91 & 36.19\\
T & 0:25:14.58 & 64:07:55.0 & -3.85 & 0.52 & 0.72 & 0.62 & -3.77 & 0.91 & 36.78\\
K & 0:25:23.89 & 64:08:39.3 & 0.18 & 0.17 & 0.73 & 0.69 & 0.20 & 0.15 & 38.73\\
L & 0:25:24.30 & 64:08:40.5 & 0.16 & -0.44 & 0.75 & 0.82 & 0.11 & -0.45 & 41.59\\
S & 0:25:13.78 & 64:08:34.4 & 4.16 & 0.58 & 0.83 & 0.84 & 4.20 & 0.15 & 42.09\\
2 & 0:25:22.44 & 64:07:32.4 & 74.85 & -4.43 & 0.82 & 0.83 & 74.05 & -11.94 & 46.09\\
\enddata
\end{deluxetable}

\subsection{Radial Velocity}
\label{sec:radvel}

For the radial-velocity measurement we first obtained well-calibrated wavelength solutions for our spectra. MAKEE performs an initial calibration of the wavelength using arcs and then refines it by cross-correlating the night-sky lines for each observation and determining minor offsets. Both science objects and radial-velocity standards were reduced in the same fashion. 

Each order of each star spectrum was then cross-correlated using the \gls{iraf} task \textsc{fxcor} \citep[][]{1979AJ.....84.1511T} with at least two other radial-velocity standards (\object{HR6349}, \object{HR6970}, and \object{HR1283}) which had been observed on the same night. We measure the radial velocity of the standards and, comparing to the canonical values \citep{1999ASPC..185..354S}, we obtain a systematic error of $\sim 1\,\kms$, which is negligible compared to the measured velocities.

The radial velocity of \starb\ was measured in the course of determining the stellar parameters of \starb\ with the stellar parameter fitting package \gls{sfit}. The \gls{sfit} result consistently gives $v_\textrm{helio} = -52\,\kms$ for different stellar parameters with an uncertainty of $\sim 2\,\kms$. 

In Table~\ref{tab:radvel} we list all of the radial velocities both in a heliocentric frame and a \gls{lsr} frame. We will be referring to the heliocentric measurements henceforth. The listed uncertainty is the standard deviation of the radial-velocity measurement of all orders added in quadrature to the error of the radial-velocity standards.

\begin{figure*}[ht!] 
   \centering
   \includegraphics[width=\textwidth]{\plotdir 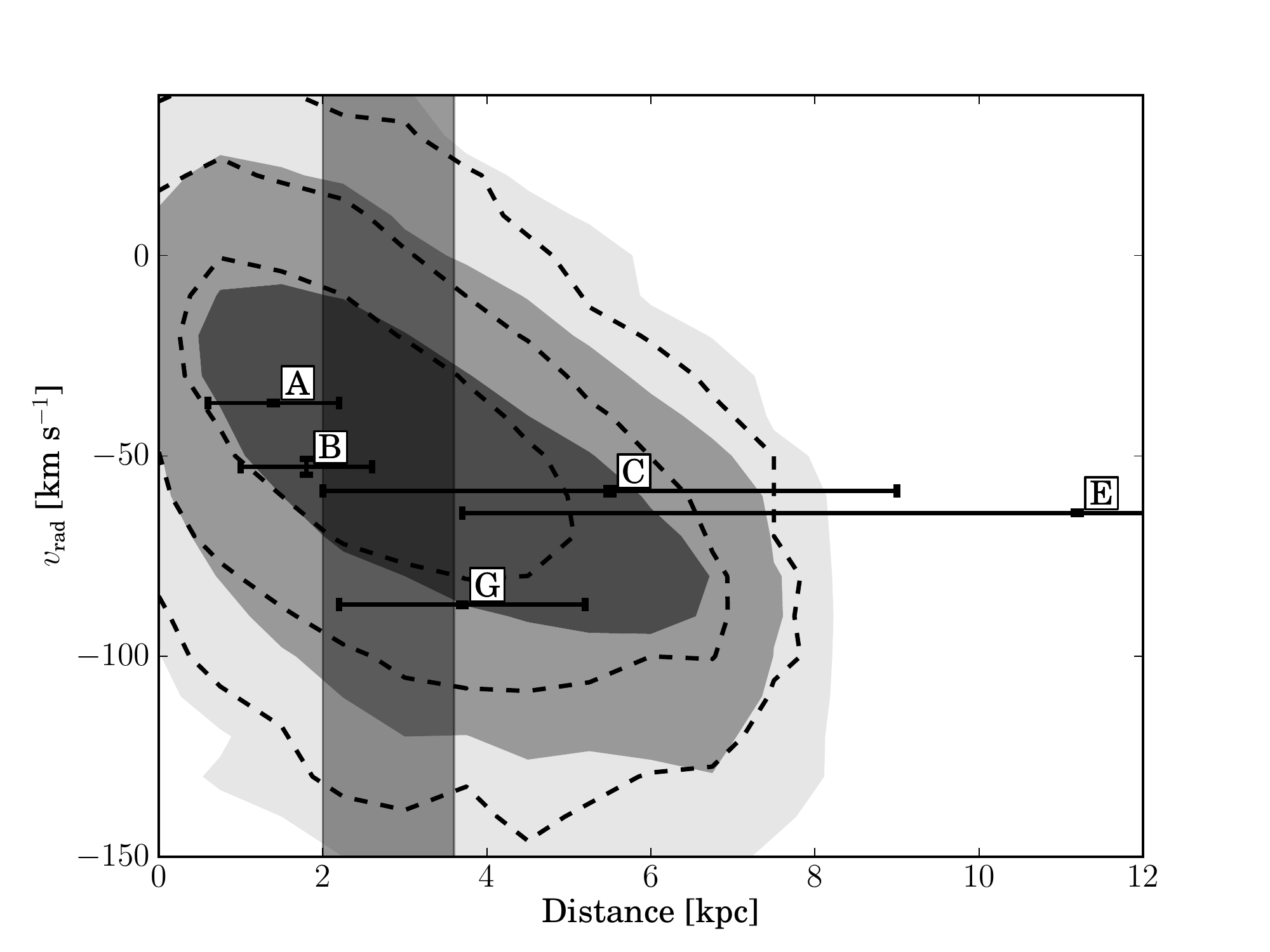} 
   \caption{The fcontours indicate $1\sigma$, $2\sigma$, and $3\sigma$ levels of the distance and radial velocity using the Besan\c{c}on Model \citep{2003A&A...409..523R} with $\sim$ 60,000 stars in the direction of SN~1572 (only including stars with $10  < V < 20$ mag and with a metallicity of [Fe/H] $> -1$ for the filled contours and [Fe/H] $>-0.2$ for the dashed contours). We have overplotted our candidate stars with error bars. One should note that the uncertainties in distance are a marginalized approximate of the error; the proper error surfaces can be seen in Figure~\ref{fig:mc_isochrone}. The vertical gray shade shows the error range for the distance to SNR~1572.}
   \label{fig:dist_vr}
\end{figure*}

In Figure \ref{fig:dist_vr} we compare the radial velocity of our sample stars to radial velocities of stars in the direction of Tycho's SNR using the Besan\c{c}on Model \citep{2003A&A...409..523R}. The distances as well as their uncertainties are taken from \S \ref{sec:distance}.  The candidates' radial velocities are all typical for their distances. Finally, we note that the measurement of \starg\ is consistent with the results of \wek\ and \gh.

\begin{deluxetable}{lccccccccc}
\tablecaption{Radial Velocities of Candidates\label{tab:radvel}}
\tablecolumns{10} 
\tablehead{ 
\colhead{Tycho} &          
\colhead{Date} &   		      
\colhead{$v_\textrm{helio}$} &
\colhead{$v_\textrm{LSR}$} &
\colhead{$\sigma$} \\%
%
\colhead{(Name)}&%
\colhead{(dd/mm/yy)}&%
\colhead{(dd:mm:ss.ss)}&%
\colhead{(\kms)}&%
\colhead{(\kms)}%

}
\startdata
\stara & 09/09/06 & $-36.79$ & $-28.50$ & 0.23   \\
\starb & 09/09/06 & $-52.70$ & $-44.41$ & $\sim 2$ \\
\starc & 11/10/06 & $-58.78$ & $-50.49$ & 0.75   \\
\stard & 11/10/06 & $-58.93$ & $-50.64$ & 0.78   \\
\stare\tablenotemark{a} & 11/10/06 & $-64.20$ & $-55.91$ & 0.27   \\
\starg & 09/09/06 & $-87.12$ & $-78.83$ & 0.25 \\
\starg & 11/10/06 & $-87.51$ & $-79.22$ & 0.78 
\enddata
\tablenotetext{a}{There seems to be a discrepancy between \rl\ and this work (measurement by \rl\ $v_\textrm{LSR}$ $-26$\,\kms), which might be a possible hint of a binary system.}
\end{deluxetable}


\subsection{Rotational Velocity}
\label{sec:rotation}
We have measured projected rotational velocities ($\vrot \sin{i}$) of all stars except \starb\ in the fashion described by \wek. We selected several unblended and strong (but not saturated) \ion{Fe}{1} lines in the stellar spectra, and added them after shifting to the same wavelength and scaling to the same \gls{ew}. This was done to improve the S/N for the faint stars as well as provide consistency throughout all stars. 

As a reference we created three synthetic spectra for each star (one broadened only with the instrumental profile, the others with the instrumental profile and a $\vrot\sin{i}$\ of 10 and 13\,\kms, respectively) with the 2010 version of \gls{moog}, using our derived temperature, surface gravity, and metallicity.  As input data to \gls{moog} we used the \citet{2004astro.ph..5087C} atmospheric models and a line list from \citet{1995KurCD..23.....K}. We then applied the same process of line selection and adding as for the lines in the observed spectra.

\begin{figure*}[ht!]
\begin{tabular}{cc}
\includegraphics[width=0.45\textwidth, trim=130 30 60 0]{\plotdir 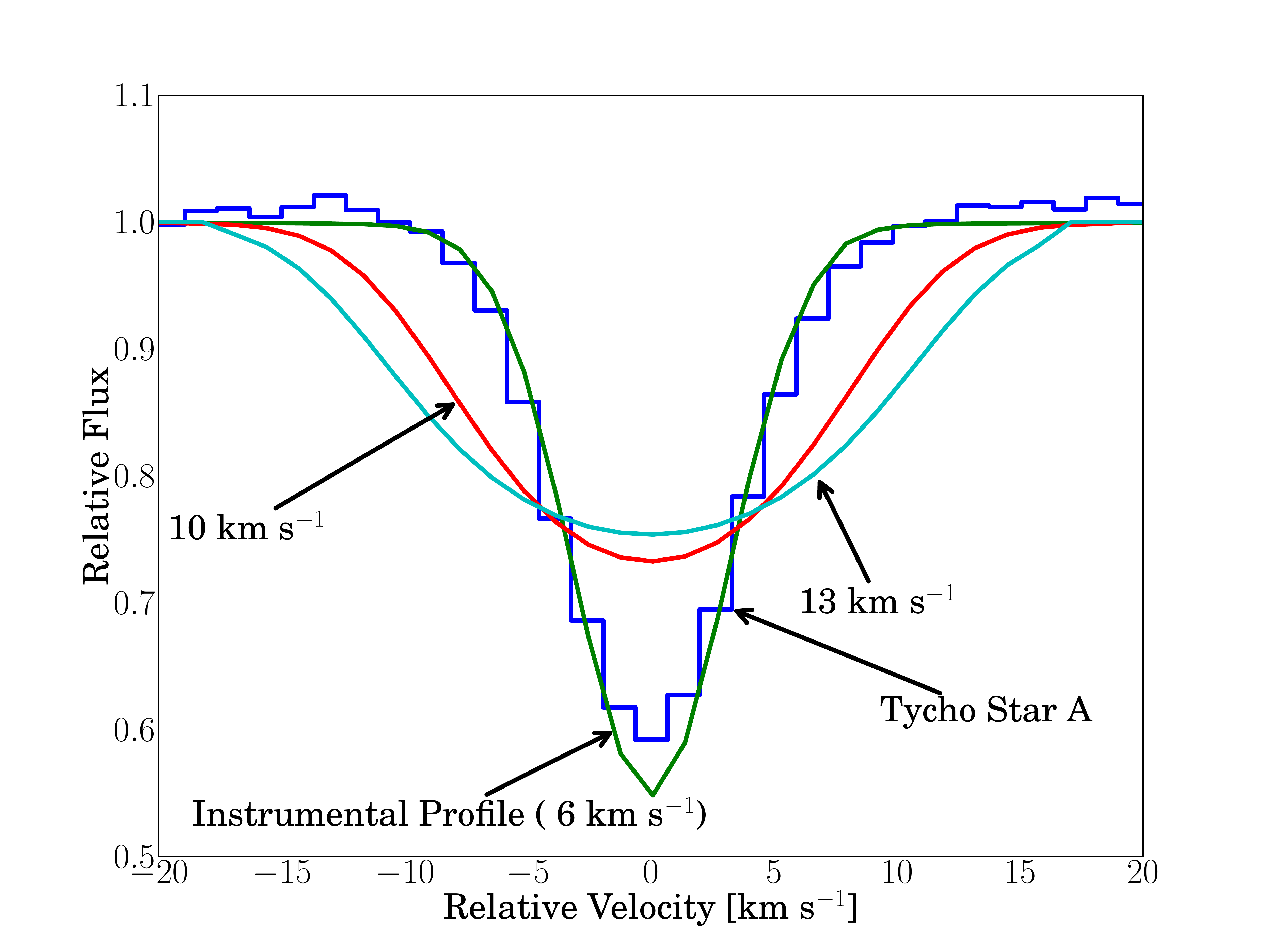} &
\includegraphics[width=0.45\textwidth, trim=130 30 60 0]{\plotdir 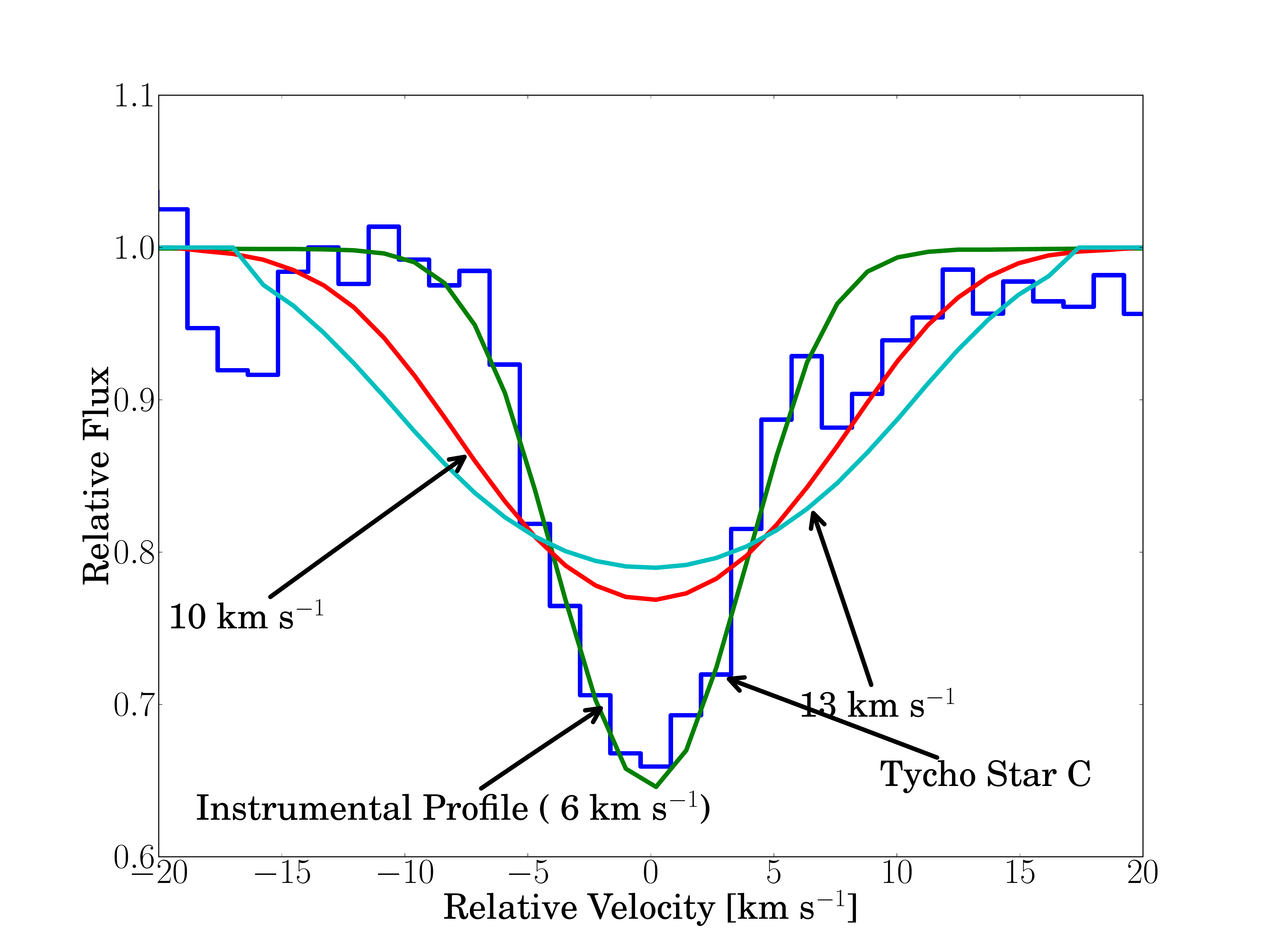} \\
\includegraphics[width=0.45\textwidth, trim=130 30 60 0]{\plotdir 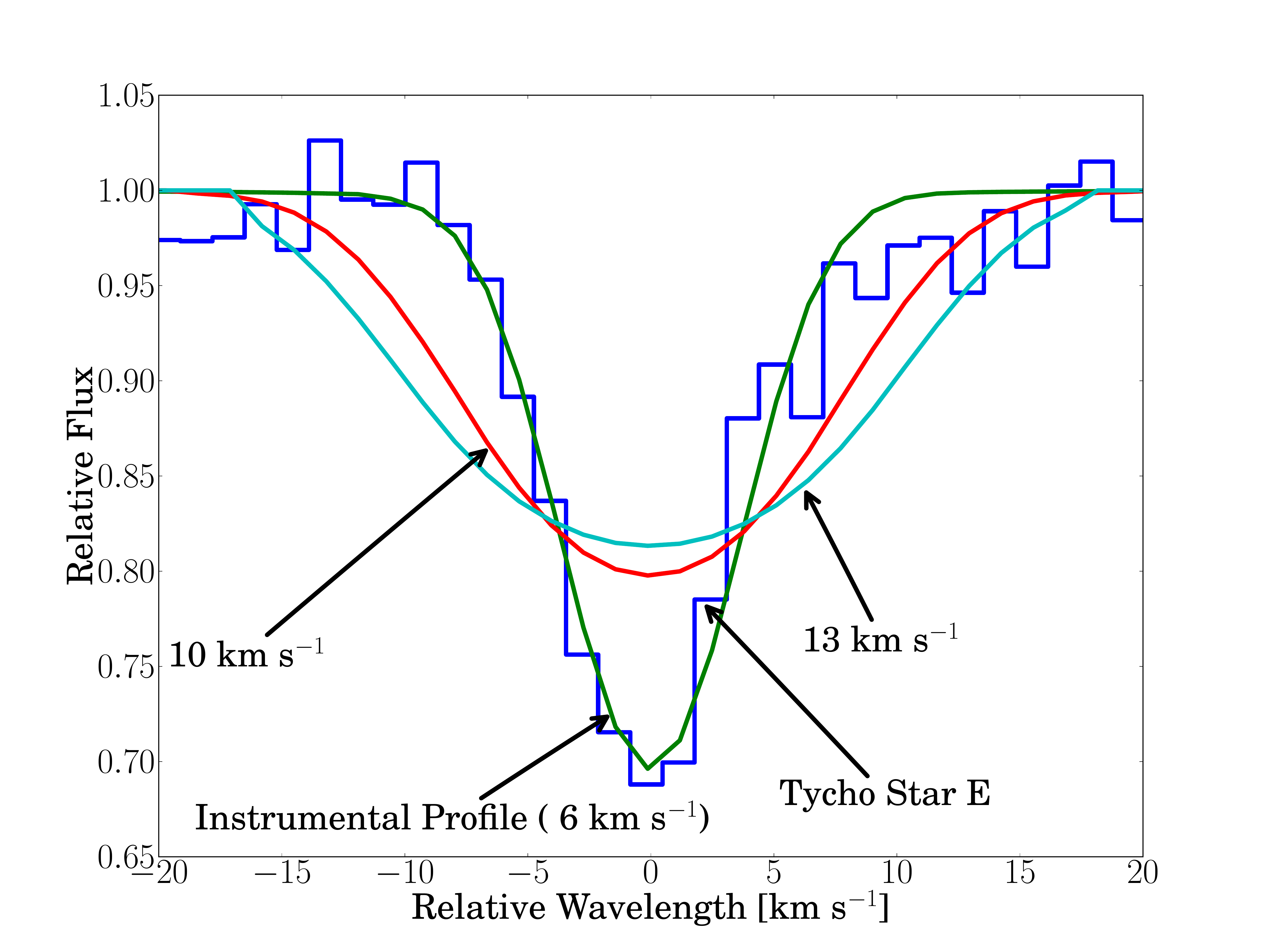} &
\includegraphics[width=0.45\textwidth, trim=130 30 60 0]{\plotdir 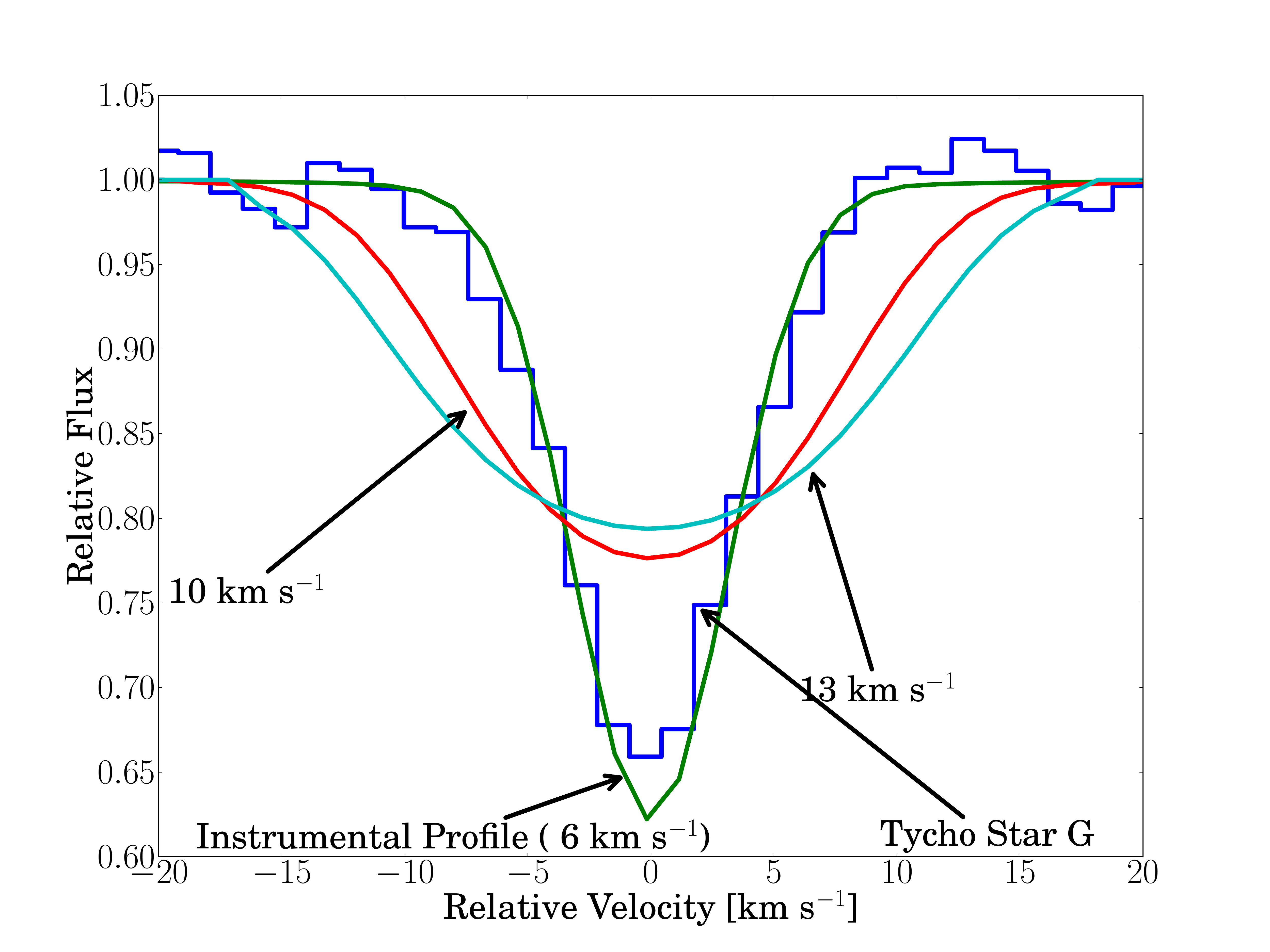} \\
\end{tabular}
\caption[Rotation measurement for all candidate stars in SN~1572]{The figures show the combination of iron line profiles after normalization to the same EW and compare them to synthetic line profiles created by \glsentryname{moog}. We convolved the synthetic lines first with a rotational kernel with three different values for rotation and then with the instrumental profile. All stars show rotation less than 6\,\kms, which is equal to the instrumental profile at this resolution. }
\label{fig:sn1572_hires:rotvel}
\end{figure*}

Figure \ref{fig:sn1572_hires:rotvel} shows the comparison between the synthetic spectra with different rotational velocities and the observed spectra. This comparison indicates that the stellar broadening (rotational, macroturbulence, etc.) is less than broadening due to the instrumental profile of 6\,\kms\ for each star. We adopt 6\,\kms\ as an upper limit to the rotation for all stars.

Due to its high temperature and rotation, we fit the rotational velocity for \starb\ with the program \gls{sfit} \citep[][described in \S \ref{sec:stellar-parameters}]{2001A&A...376..497J}  as part of the overall fit for this star's stellar parameters.  We find $\vrot \sin{i}=171^{+16}_{-33}$\,\kms. While \starb's rotation is very high compared to the other candidate stars,  for stars of this temperature and surface gravity a high rotation is not unusual. In summary, other than \starb, none of the stars show rotation which is measurable at this resolution.

\subsection{Stellar Parameters and Chemical Abundances}
\label{sec:stellar-parameters}
The stellar parameters are presented in Table~\ref{tab:param} and were determined using 
a traditional spectroscopic approach. 
Due to its high temperature, we measure the stellar parameters for \starb\  
by direct comparison to models in a separate procedure described later in this subsection. 

The first step in the spectroscopic analysis was to rectify the continuum. For each order, we 
fit the continuum, by eye, using a low-order polynomial function within the \textsc{continuum} 
task in \gls{iraf}. To help identify continuum regions in the program stars, we made use of the 
Arcturus and solar spectra \citep{2000IAUJD...1E..26H}. Consideration of the moderate 
S/N was a concern. For example, at these values of the S/N,  
we were mindful of not fitting the continuum to the highest points since it is 
likely that these values are noise rather than true continuum regions. 

Next, equivalent widths (EWs) for a set of Fe and Ni 
lines were measured using routines in \gls{iraf}. 
The $\log\,{gf}$ values for the \fei\ lines were from 
the laboratory measurements by the Oxford group 
\citep[e.g., ][ and references therein]{1979MNRAS.186..633B,1979MNRAS.186..657B,1980MNRAS.191..445B,1986MNRAS.220..549B,1995A&A...296..217B} and the \feii\ lines were from the measurements by \citet{1991A&A...249..539B}. 
For Ni, the $\log\,{gf}$ values were taken from the compilation 
by \citet[][henceforth Reddy03]{2003MNRAS.340..304R} 
and \citet[][henceforth RC02]{2002AJ....123.3277R}. 
While these EW measurement 
routines employ Gaussian fits in a semi-automated manner, we 
emphasize that all EWs were visually checked on at least two occasions. 
We also required that lines have an EW of at least 10\,m\AA\ to avoid measuring noise 
and less than $\sim 150$\,m\AA\ to avoid saturated lines with non-Gaussian profiles 
that may lie on the flat part of the curve-of-growth. 
Table~\ref{tab:ew} shows the EWs measured for the program stars. 
Missing values indicate that the line was not detected or that no reliable measurement 
could be obtained. 
In the following subsection, we consider in more detail the uncertainties that 
arise from continuum placement and EW errors. 

We used the 2011 version \citep{2011AJ....141..175S} of the 
local thermodynamic equilibrium (LTE) stellar line analysis program 
MOOG \citep{1973ApJ...184..839S} 
and LTE model atmospheres from the \citet{2003IAUS..210P.A20C} grid to derive an 
abundance for a given line. The effective temperature, \teff, was adjusted until the 
abundances from \fei\ lines displayed no trend as a function of 
lower excitation potential, $\chi$. 
The surface gravity, $\log\,g$, 
was adjusted until the abundances from \fei\ and \feii\ lines were 
in agreement. The microturbulent velocity, $\xi_t$, was adjusted until there 
was no trend between the abundances from the \fei\ lines and the reduced 
EW, log\,(EW/$\lambda$). This process was iterated until self-consistent 
stellar parameters were obtained for each star. 

In our analysis, we explored stellar parameters at discrete values. 
For effective temperature, we considered values at every 25\,K (e.g., 6000, 6025\,K, etc.); 
for surface gravity, we considered values at every 0.05\,dex (e.g., 4.00, 4.05\,dex, etc.);
and for $\xi_t$, we considered values at every 0.01\,\kms\ (e.g., 1.70, 
1.71\,\kms, etc.). We assumed that excitation equilibrium was satisfied when 
the slope between $\log\,\epsilon$(\fei) and lower excitation potential ($\chi$) was $\le 0.004$. 
We assumed that ionization equilibrium was achieved when  $\log\,{\epsilon}($\fei$) -
\log\,\epsilon($\feii$) \le 0.02$\,dex. The microturbulent velocity was set when 
the slope between $\log\,\epsilon($\fei$)$ and reduced EW (log\,EW/$\lambda$) was $\le 0.004$. 
We found a unique solution for all program stars. 
We estimate that the internal uncertainties are 
typically \teff\ $\pm$ 100\,K, log\,$g$ $\pm$ 0.3\,dex, and $\xi_t$ $\pm$ 0.3 \kms. 
For further details regarding the derivation of stellar parameters, see \citet{2008ApJ...673..854Y}.

The final iron measurements are the average of \fei\ and \feii, weighted 
by the number of lines measured for each species. 
We adopted the solar abundances of \citet{2009ARA&A..47..481A}. 

In addition, we measured element abundance ratios for Ni via EW analysis 
and Li (only for \starg) via spectrum synthesis (see Figure ~\ref{fig:li_synth}). 
For the Li spectrum synthesis, we used 
the \citet{2002MNRAS.335.1005R} line list in combination with 
MOOG and the \citet{2003IAUS..210P.A20C} model atmospheres. A non-LTE (NLTE) analysis \citep{2009A&A...503..541L} of the Li abundances ($A$(Li)$_{\rm NLTE}$ = 2.45) yields nearly the same result as the LTE abundance ($A$(Li) = 2.46). Abundances are presented in Table~\ref{tab:parvar}. 
Tycho-B's abundances are not presented in the table as they were measured in a different fashion.

\begin{figure*}[ht!] 
   \centering
   \includegraphics[width=1\textwidth, trim=0 0 0cm 0, clip]{\plotdir 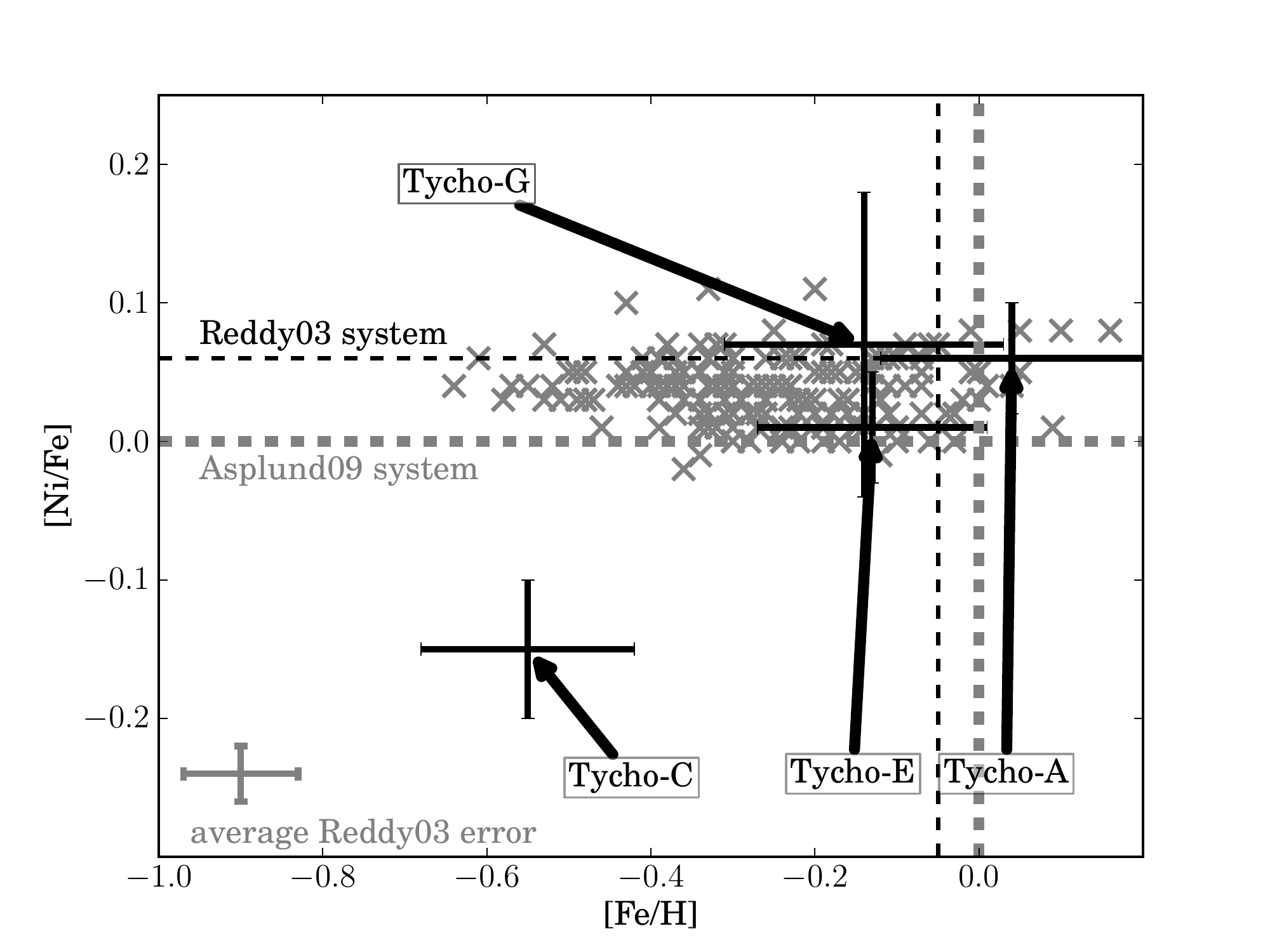} 
   \caption[Comparison of nickel and iron abundance measurements of stars in SN~1572]{The background gray error bars are F- and G-dwarf abundances from \citet{2005A&A...433..185B}. All candidate stars are consistent with that distribution. \starc\ can be seen as an outlier, but it is a K-giant and its class is not represented in the underlying F- and G-dwarf distribution.}
   \label{fig:bensby05}
\end{figure*}

In summary, the inferred metallicities for all candidates show that the candidates are of roughly solar metallicity with the exception of the metal-poor \starc. The range of metallicities spanned by the program stars is compatible with membership in the thin disk. Based on metallicity alone, we do not regard any of the program stars to be unusually metal-poor or metal-rich.  Additionally, we have compared the [Ni/Fe] abundance ratio to a well-calibrated set of F- and G-dwarf abundances \citep{2005A&A...433..185B}, which we calibrated to the solar abundances of \citet{2009ARA&A..47..481A}. Figure~\ref{fig:bensby05} shows that all program stars are consistent with stars of similar metallicity. We do note that \starc\ is a marginal outlier (perhaps $1\sigma$) with a low [Ni/Fe] abundance ratio, but do not regard this to be significant. To avoid selection effects we compared \starc\ to a sample of giant stars \citep{2007AJ....133.2464L}, which gives a similar result as the comparison with \citet{2005A&A...433..185B}.

Because \starb\ has a temperature  greater than 9000\,K and is rapidly rotating, the process described above cannot be used to measure stellar parameters. Instead we used the program \gls{sfit} to match the \gls{hires} spectrum to a grid of model spectra. To determine the stellar parameters for \starb\ we have used a model grid with $\feh= -1.0$, $8000 < \teff <$ 16,000\,K, and $7 < \log\,{g} < 2$. This low metallicity is suggested by the very weak Ca~II K line and \ion{Mg}{2} lines, but it is hard to measure. We cannot measure helium directly in this spectrum and thus adopt $N$(He) = 0.1, as this is empirically a very common helium abundance in stars.

This analysis resulted in $\teff = 10,000^{+ 400}_{-200}\,\textrm{K}$, $\log\,{g} = 3.67$ with slope  $\partial \log\,g/\partial \teff = 0.27/500\,\textrm{K}^{-1}$, and rotational velocity $\vrot \sin{i} = 171$\,\kms\ with slope $\partial \vrot \sin{i} / \partial \teff = -41/500\,\kms\,\textrm{K}^{-1}$. From qualitative analysis this object seems metal poor (e.g., in comparison to stars of similar stellar parameters but solar metallicity), but its high rotation and temperature make it hard to determine this parameter precisely. For the present, we assume [Fe/H] = $-1.0$ unless otherwise noted.

In addition, using the high-resolution spectrum, we measured the \glspl{ew} of several lines predicted to be strong in the \gls{vald}. The abundances were deduced from the \glspl{ew} using a model atmosphere having $T_{\rm eff} =$ 10,000\,K, $\log\,{g} = 3.67$, and [Fe/H] = $-1.0$ (see Table~\ref{tab:starb-abund}).

One caveat regarding these abundances is the use of \glspl{ew} from 
single lines with large rotational broadening, since the effect of blending 
with nearby weak lines cannot be taken into account. A second is that these 
abundances invariably rely on the strongest lines, which are precisely those 
most susceptible to departures from LTE.
Nevertheless, they do confirm the earlier impression that the star is 
metal-poor, and justify the adoption of \feh\ = $-1.0 \pm 0.4$.

\begin{deluxetable}{lcccccc}
\tablecaption{Tycho-B abundances\label{tab:starb-abund}}
\tablecolumns{7} 
\tablehead{ 
\colhead{Ion} &
\colhead{$\lambda$}&
\colhead{$W_\lambda$} &  
\colhead{$\epsilon$}&
\colhead{$[X/H]$} &
\colhead{$\frac{\partial \epsilon}{\partial \log\,g}$}  &
\colhead{$\frac{\partial \epsilon}{\partial \teff}$}\\
\colhead{designation} &
\colhead{(\AA)} & 
\colhead{(\AA)} & 
\colhead{(dex)} & 
\colhead{(dex)}&
\colhead{} &
\colhead{(K$^{-1}$)}\\
}

\startdata
\ion{Mg}{2} & 4481.13+4481.33 & $220\pm15$ & $6.18\pm.08$ & -1.40&0.08&$8\times10^{-5}$ \\ 
\ion{Si}{2} & 6347.1 & $140\pm5$ & $6.96\pm.18$ & -0.59&-0.02&$1\times10^{-4}$\\
\ion{O}{1} & 7771.9+7774.2+7775.4 & $460\pm30$ & $8.43\pm.10$ & -0.58 &0.24&$-4\times10^{-5}$
\enddata
\end{deluxetable}

As a second approach to determine the stellar parameters of \starb\ we used the low-resolution spectrum obtained with \gls{lris}.  The observation range of \gls{lris} was chosen to be centered around the Balmer jump, as this feature is sensitive to the surface gravity \citep{2007PASP..119..605B}. We fitted the spectrum to a grid of model spectra \citep[]{2005A&A...442.1127M} using a spectrum-fitting tool  described below. The final grid we used covered $\log\,{g}$ from 3.5 to 4.5 in steps of 0.5 and \gls{teff} from 9000 to 12,000\,K in steps of 500\,K. In addition, we expanded the grid by reddening the spectra with the \textsc{pysynphot}\footnote{The \textsc{pysynphot} package is a product of the Space Telescope Science Institute, which is operated by AURA for NASA.} package. We also added diffuse interstellar bands  \citep{1937PASP...49..224B, 1966ZA.....64..512H, 1967IAUS...31...85H, 1975ApJ...196..129H, 1995ARA&A..33...19H, 1994dib..nasa...31H, 1994A&AS..106...39J, 1958ApJ...128...57W} to the synthetic spectra, scaled with reddening. The included $E(B-V)$ ranged from 0.5 to 1.3 mag in steps of 0.2. We assumed a rotation of 171\,\kms\ in the grid  (see \S \ref{sec:rotation}).

\begin{figure*}[ht!]

\includegraphics[width=1.\textwidth]{\plotdir 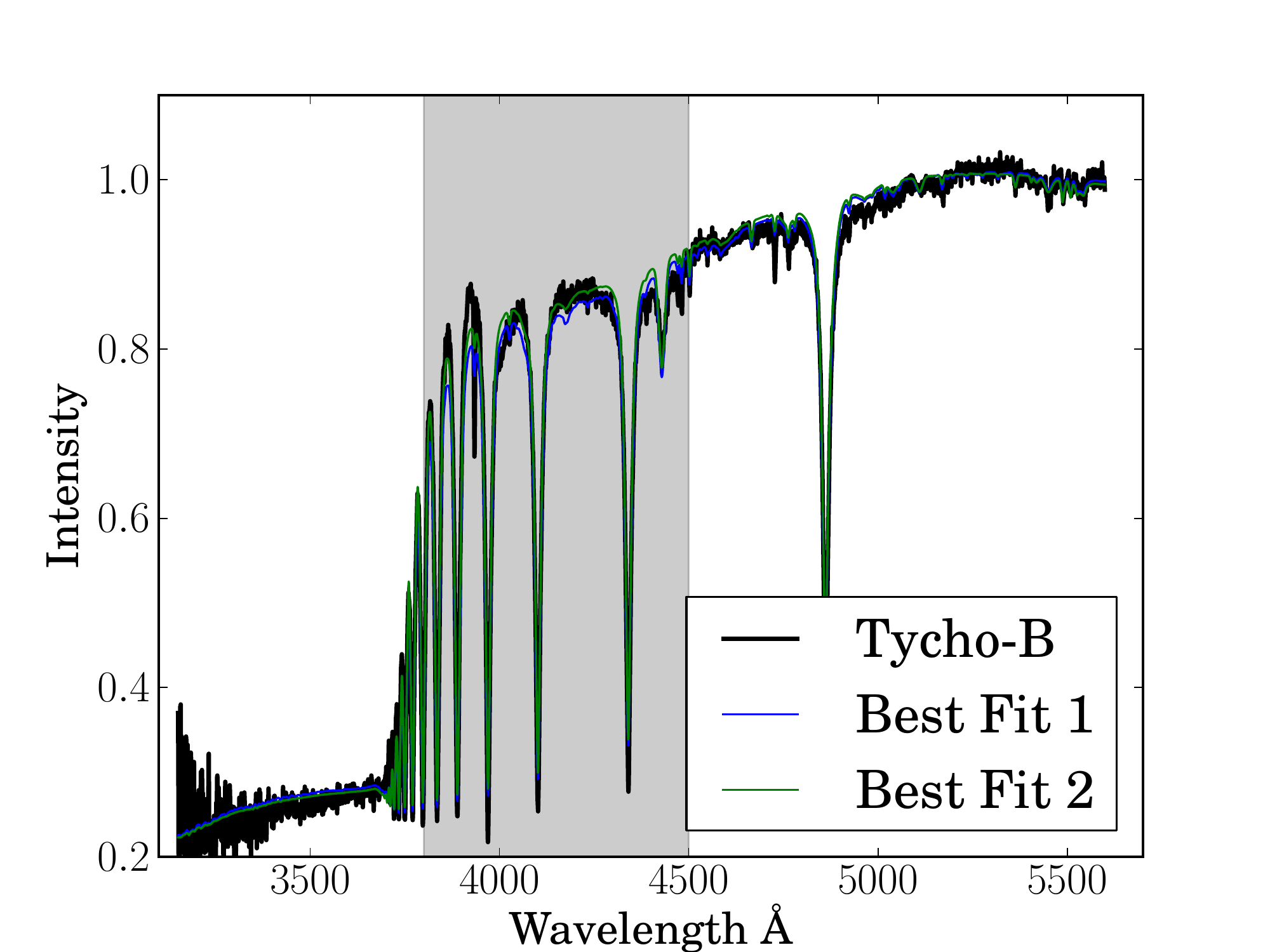} 

\caption[Fit of low-resolution spectrum of Tycho-B]{The plot shows the normalized spectrum of \starb\ with the fit which excluded the spectral region 3800--4500\,\AA\ (Best Fit 1) and the fit with the problematic region (Best Fit 2). The region is marked with a grey shade.  }
\label{fig:starb_spec_comp}
\end{figure*}

We used $\chi^2$ as a figure of merit in our fitting procedure. To find the best fit for \starb\ we used the \gls{migrad} algorithm provided by \gls{minuit} and linearly interpolated between the grid points using \textsc{LinearNDInterpolator} provided by the \gls{scipy} package. The fit of \starb\ results in \glssymbol{teff} = 10,570\,K, $\log\,g = 4.05$, \feh\ = $-1.1$, and $E(B-V) = 0.85$\,mag. The model fits the synthetic spectrum poorly  in the wavelength region 3800--4280\,\AA\ (see Figure \ref{fig:starb_spec_comp}). The adopted mixing-length parameter in one-dimensional (1D) model atmospheres, used to construct the spectral grid, influences the fluxes in that region and affects the hydrogen line profiles. \citet{2002A&A...392..619H} and others show that a mixing length of 0.5, rather than 1.25 as used in the Kurucz/Munari grid, better fits the violet fluxes and the H line profiles. Spectra using a mixing-length parameter of 0.5 are brighter in the ultraviolet, and the H$\delta$, H$\gamma$, and H$\beta$ profiles give the same \gls{teff} as the \gls{halpha} profiles. We have chosen, however, to fit the spectrum and ignore the problematic spectral region (3800--4280\,\AA) to avoid a systematic error. This yields $\teff = 10,722$\,K, $\log\,g=4.13$, $\feh\ = -1.1$, and $E(B-V) = 0.86$\,mag. The differences are indicative of the size of systematic errors in the model fits. We adopt the fit excluding the problematic wavelength region in the subsequent analysis. Exploring the complex search space, we estimate the uncertainties to be $\Delta\teff = 200$\,K, $\Delta\log\,g = 0.3$, and $\Delta\feh = 0.5$, and we note that the parameters are correlated.
\begin{deluxetable}{ccccr}
\tabletypesize{\scriptsize}
\tablecolumns{5} 
\tablewidth{0pc} 
\tablecaption{Stellar Parameters\label{tab:param}}
\tablehead{ 
\colhead{Tycho} &					
\colhead{\teff} &                                          
\colhead{\logg} &                 
\colhead{$\xi_t$} &                 
\colhead{[Fe/H]}  \\     
\colhead{} &
\colhead{(K)} &
\colhead{(dex)} &
\colhead{(\kms)} &
\colhead{(dex)} 
}
\startdata 
A & 4975 & 2.9 & 1.20 &    0.04 \\

B & 10722 & 4.13 & 2 & -1.1 \\

C & 4950 & 2.9 & 2.14 & $-$0.55 \\
E & 5825 & 3.4 & 1.82 & $-$0.13 \\
G & 6025 & 4.0 & 1.24 & $-$0.13 \\
\enddata
\end{deluxetable}

\begin{deluxetable}{lccrccccrrcccc}
\tabletypesize{\scriptsize}
\tablecolumns{14} 
\tablewidth{0pc} 
\tablecaption{Chemical Abundances and Error Estimates 
for the Program Stars. \label{tab:parvar}}
\tablehead{ 
\colhead{Species} & 			
\colhead{$\log \epsilon (X)_\odot$} & 	
\colhead{$\log \epsilon (X)$} & 	
\colhead{[X/H]} &			
\colhead{[X/Fe]} &			
\colhead{$\sigma$} & 			
\colhead{N$_{\rm lines}$} & 		
\colhead{$\Delta_\sigma$} & 		
\colhead{$\Delta_{T_{\rm eff}}$} &		
\colhead{$\Delta_{\log g}$} &		
\colhead{$\Delta_\xi$} &		
\colhead{$\Delta_{\rm [m/H]}$} &	
\colhead{$\Delta{\rm [X/H]}$} &		
\colhead{$\Delta{\rm [X/Fe]}$} \\	
\colhead{(1)} &
\colhead{(2)} &
\colhead{(3)} &
\colhead{(4)} &
\colhead{(5)} &
\colhead{(6)} &
\colhead{(7)} &
\colhead{(8)} &
\colhead{(9)} &
\colhead{(10)} &
\colhead{(11)} &
\colhead{(12)} &
\colhead{(13)} &
\colhead{(14)} 
}
\startdata 
 \noalign{\vskip +0.5ex}
 \multicolumn{14}{c}{Star A} \cr
 \noalign{\vskip  .8ex}%
 \hline
 \noalign{\vskip-2ex}\\
Fe I  & 7.50   &     7.54 &0.04  & \ldots & 0.16 & 40 & 0.03 & $-$0.08 & $-$0.01 &    0.14 &    0.02 &    0.16 & \ldots \\
Fe II & 7.50   &   7.54 & 0.04 &   \ldots & 0.15 & 14 & 0.04 &    0.07 & $-$0.16 &    0.12 &    0.06 &    0.22 & \ldots \\
Ni I  & 6.22   &      6.33 & 0.10 &  0.06 & 0.08 & 37 & 0.01 & $-$0.05 & $-$0.04 &    0.11 &    0.03 &    0.13 &   0.04 \\
\cutinhead{Star C}
Fe I  & 7.50   &  6.95& $-$0.55 & \ldots & 0.23 & 36 & 0.04 & $-$0.10 & $-$0.02 &    0.07 &    0.01 &    0.13 & \ldots \\
Fe II & 7.50   &  6.96& $-$0.54 & \ldots & 0.27 &  5 & 0.12 &    0.05 & $-$0.16 &    0.05 &    0.06 &    0.22 & \ldots \\
Ni I  & 6.22   &  5.52 & $-$0.70 &$-$0.15 & 0.21 & 20 & 0.05 & $-$0.06 & $-$0.04 &    0.07 &    0.03 &    0.10 &   0.05 \\
\cutinhead{Star E}
Fe I  & 7.50    &7.37 &  $-$0.13 &  \ldots & 0.22 & 40 & 0.03 & $-$0.09 &    0.02 &    0.11 &    0.00 &    0.14 & \ldots \\
Fe II & 7.50     &7.38 & $-$0.12 &\ldots & 0.16 &  7 & 0.06 &    0.03 & $-$0.14 &    0.08 &    0.03 &    0.18 & \ldots \\
Ni I  & 6.22   &     6.1 & $-$0.12 & 0.01 & 0.17 &  8 & 0.06 & $-$0.07 &    0.00 &    0.09 &    0.00 &    0.11 &   0.04 \\
\cutinhead{Star G}
Fe I  & 7.50  & 7.37 & $-$0.13 &  \ldots & 0.18 & 69 & 0.02 & $-$0.09 &    0.03 &    0.09 &    0.01 &    0.13 & \ldots \\
Fe II & 7.50   &7.35& $-$0.15 &  \ldots & 0.18 & 16 & 0.05 &    0.01 & $-$0.12 &    0.07 &    0.03 &    0.15 & \ldots \\
Ni I  & 6.22   &6.16 & $-$0.06 &    0.07 & 0.14 & 18 & 0.03 & $-$0.07 &    0.01 &    0.05 &    0.01 &    0.09 &   0.04 \\
Li   & \ldots & 2.46& \ldots & \ldots & \ldots & 1 & 0.05\tablenotemark{a} & $-$0.07 & 0.01 & 0.00 & 0.02 & 0.09 & \ldots \\ 
\enddata

\tablenotetext{a}{This represents the uncertainty in the fitting of the line.}

\end{deluxetable}


\subsection{Tycho-G: A Detailed Comparison with GH09} 
\label{sec:tychog_comp}
GH09 suggested that \starg\ is a plausible donor star, with the primary evidence 
consisting of an unusually high Ni abundance and a high 
space velocity (radial velocity and proper motion). 
In this subsection, we focus on the Ni abundance, and we refer the reader to 
\S 3.1, 3.2, and 4 on the proper motion and radial velocity.
 
The measured values are
[Ni/Fe] = 0.16 $\pm$ 0.04 and 0.07 $\pm$ 0.04 for GH09 and this study, respectively, from the same HIRES spectra. 
The magnitude of the difference is 0.09 dex, and it is 
significant at the $\sim 1.5\sigma$ level. 
While our [Ni/Fe] ratio in \starg\ is lower than that measured by 
GH09, our value does not represent a substantial revision given the 
measurement uncertainties involved. Nevertheless, our [Ni/Fe] 
measurement and comparison with the literature do not support an unusually high 
Ni abundance, and we conclude that \starg\ does not show any obvious chemical signature 
that one may seek to attribute to a supernova companion star. 
In order to identify the origin of the difference in [Ni/Fe] ratios, 
we now compare our stellar parameters and chemical abundances to those of GH09.

Both studies determined stellar parameters and chemical 
abundances in a similar manner, from 
a standard spectroscopic EW analysis using 1D LTE Kurucz model atmospheres and 
the MOOG stellar line analysis software. Our analysis employed more recent 
versions of both tools. The first test we can perform is to use the GH09 line list 
and stellar parameters but with our tools --- namely, 
the 2011 version of MOOG \citep{2011AJ....141..175S, 1973ApJ...184..839S} 
and the \citet{2003IAUS..210P.A20C} model atmospheres. 
Adopting this approach, we obtain 
$\log\,\epsilon$(\fei) = 7.38 ($\sigma$ = 0.13), 
$\log\,\epsilon$(\feii) = 7.42 ($\sigma$ = 0.10), and 
$\log\,\epsilon$(Ni I) = 6.33 ($\sigma$ = 0.19). 
These values are in very good agreement with those of GH09, who obtained 
$\log\,\epsilon$(\fei) = 7.42 ($\sigma$ = 0.12), 
$\log\,\epsilon$(\feii) = 7.42 ($\sigma$ = 0.10), and 
$\log\,\epsilon$(Ni I) = 6.36 ($\sigma$ = 0.19). 
Thus, we argue that any abundance differences (for Fe and Ni) between the two studies, 
exceeding the $\sim$0.04 dex level, cannot be attributed to differences 
in the model-atmosphere grid and/or line-analysis software. 

Our stellar parameters (\teff = $5900 \pm 100$K, $\log\,g = 3.85 
\pm 0.30$, [Fe/H] = $-0.05 \pm 0.09$) are in good agreement with 
those of GH09 (\teff = $6000 \pm 100$, $\log\,g = 4.00 \pm 0.30$, 
[Fe/H] = $-0.13 \pm 0.13$). The second test we can perform is to 
determine chemical abundances using (i) the GH09 stellar parameters but with 
our line list and (ii) our stellar parameters and line list. 
On comparing case (ii) minus case (i), we find 
$\Delta\log\,\epsilon$(\fei) = 0.10, 
$\Delta\log\,\epsilon$(\feii) = 0.02, and 
$\Delta\log\,\epsilon$(Ni I) = 0.08.  
Adopting the same solar abundances and method for determining the 
average [Fe/H] value (average of \fei\ and \feii\ weighted by the number of lines)
as in the present study, we find $\Delta$[Ni/Fe] = 0.00. 
We argue that while there are abundance differences for 
$\log\,\epsilon(X)$ at the $\sim 0.10$ dex level, the [Ni/Fe] ratio 
remains unchanged, and therefore 
any differences in the [Ni/Fe] ratio between the two studies 
cannot be attributed to differences in the adopted stellar parameters. 

The solar abundances for Fe and Ni differ between the two studies. 
\gh\ adopt 7.47 and 6.25 for Fe and Ni, respectively, while we 
use 7.50 and 6.22  \citep[from ][]{2009ARA&A..47..481A}. Had we used the GH09 solar abundances, 
we would have obtained a ratio [Ni/Fe] = 0.01. Therefore, 
the different solar abundances adopted by the two studies only serve to 
decrease the discrepancy in the [Ni/Fe] ratio --- that is, any difference 
in [Ni/Fe] cannot be attributed to the solar abundances. 

The next series of comparisons we can perform concern the line lists. 
We measured Fe and Ni abundances using 
the GH09 line list but with our stellar parameters and find 
$\log\,\epsilon$(\fei) = 7.42 ($\sigma$ = 0.12), 
$\log\,\epsilon$(\feii) = 7.42 ($\sigma$ = 0.10), and 
$\log\,\epsilon$(Ni I) = 6.36 ($\sigma$ = 0.19). 
Table~\ref{tab:ni_comparison} gives a comparison of all tests performed.

Adopting the same approach as before, regarding the solar abundances and 
metallicity, yields a ratio [Ni/Fe] = 0.22, a value 
that exceeds both our measurement and that of GH09. 
We therefore speculate that the difference in [Ni/Fe] between the two
studies is driven primarily by differences in the line list. 
In particular, we note that while the \fei\ and \feii\ abundances 
are in fair agreement with our value and GH09, it is the Ni abundance, 
$\log\,\epsilon$(Ni), that shows a large difference between the two studies: 
6.16 $\pm$ 0.09 and 6.33 $\pm$ 0.10 for this study and GH09, 
respectively. Although the magnitude of this difference may appear large, 
0.17 dex, it is significant only at the $\sim 1.3\sigma$ level. 

On comparing the line lists between the two studies, we find 
3, 2, and 8 lines in common for \fei, \feii, and Ni, respectively. 
For these three species, the $\log\,{gf}$ values are on the same scale with 
differences (this study minus GH09) of $-$0.04 ($\sigma$ = 0.07), 
$-$0.03 ($\sigma$ = 0.04), and $-$0.01 ($\sigma$ = 0.03)  
for \fei, \feii, and Ni, respectively. Although the comparison sample is small, 
there is no clear evidence for any large systematic difference in $\log\,{gf}$ values 
that could explain the differing $\log\,\epsilon$(Ni) or [Ni/Fe] values. 

For the lines in common, our EWs are, on average, lower than those of GH09 
by 5.7\,m\AA\ ($\sigma$ = 8.0\,m\AA), 5.6\,m\AA\ ($\sigma$ = 5.4\,m\AA), and 12.7\,m\AA\ 
($\sigma$ = 6.9\,m\AA) for \fei, \feii, and Ni, respectively. 
The most intriguing aspect of this comparison is that the Ni lines 
show the greatest discrepancy. 
In light of the EW differences for \fei\ and \feii, we may naively have 
expected the Ni EWs to show an offset of $\sim$6\,m\AA\ rather than 
a 12.7\,m\AA\ offset. Indeed, differences in the Ni EWs appear 
to be the primary reason for the difference in the derived Ni abundances between the two studies. 

In Figure~\ref{fig:ew_comp}, we plot our EWs and 
the GH09 EWs, for the 8 Ni lines 
in common. To estimate the uncertainties in our EWs, 
we use the \citet{1988IAUS..132..345C} formula 
which considers the measurement uncertainty due to the 
line strength, S/N, 
and spectral resolution. Uncertainty in the 
continuum placement is {\it not} included in the \citet{1988IAUS..132..345C} 
formula. 

As noted in the 
previous subsection, we regard continuum placement to be an additional 
source of uncertainty in the EW measurements. To quantify this uncertainty, 
we use the DAOSPEC program which fits the 
continuum and measures EWs \citep{2008PASP..120.1332S}. 
Using DAOSPEC, we remeasure the Ni EWs using four different continuum 
fitting criteria: (i) adopting our continuum placement, and 
using a (ii) third-order, (iii) fifth-order, and (iv) ninth-order polynomial to 
refit our continuum-rectified 
spectra. For a given line, we compute the dispersion in the EW measurements 
from the four different methods for continuum fitting 
and adopt this value as being representative of the EW uncertainties due to 
continuum rectification. We then add this value, in quadrature, to the uncertainty 
using the \citet{1988IAUS..132..345C} value, noting that the latter value dominates 
the total EW error budget (see Table~\ref{tab:err_g}).

To establish whether these EW uncertainties are valid, we first identify the set of 
Ni EWs which produce our mean [Ni/Fe] ratio. That is, every line in this 
set of ``ideal'' EWs produces $\log\,\epsilon$(Ni) = 6.16, i.e., 
[Ni/Fe] = 0.07. We then added to each of these ideal EWs a 
random number drawn from a normal distribution of width corresponding to our 
estimate of the EW uncertainty. We repeated this process for each Ni line, 
computed Ni abundances for this new set of lines, and measured the 
abundance dispersion. We repeated this process for 1,000 new random samples. 
The average dispersion in Ni abundance is 0.17 dex ($\sigma$ = 0.06 dex), and 
this average value agrees well with our observed dispersion of 0.14 dex. 
Therefore, we are confident that our EW measurement uncertainties are realistic, 
since this Monte Carlo analysis verifies that these uncertainties 
reproduce our observed abundance dispersion. 

An additional test is to measure EWs from our spectra for all Fe and Ni 
lines measured by GH09. As with our EWs, all lines were manually checked. 
For \fei, we measured 27 lines and found a mean difference (this study minus GH09)
of $-$1.9\,m\AA\ $\pm$ 1.2 ($\sigma$ = 6.0). 
For \feii, we measured 8 lines and found a mean difference  
of $-$4.6\,m\AA\ $\pm$ 2.8 ($\sigma$ = 7.8). 
For Ni, we measured 18 lines and found a mean difference
of $-$8.7\,m\AA\ $\pm$ 2.0 ($\sigma$ = 8.4). 
This comparison confirms that our EWs are systematically lower than those of GH09 
and that the Ni lines, in particular, show the largest discrepancy. 
Indeed, the average difference in Ni EWs is 
4 times larger than the average difference in \fei\ EWs. 
While continuum normalization could potentially explain these differences, these 
Ni lines lie in spectral regions similar to those of the Fe lines, so we would 
expect the differences in EWs for Fe and Ni to behave similarly. 

We note in our line selection that we reject 
5, 2, and 4 lines of \fei, \feii, and Ni (respectively) that were measured by GH09. 
These lines were in our opinion blended and/or in regions where the local continuum 
was poorly defined. 


We return now to the eight Ni lines in common, noting that 
(i) for 7 of the 8 lines, our EWs are smaller than those of GH09, 
(ii) for 7 of the 8 lines, the difference in EWs exceeds 1$\sigma$, and 
for all 7 lines, the difference shows the same ``sign,'' and 
(iii) for 4 of the 8 lines, the difference in EWs exceeds 2$\sigma$, and 
for all 4 lines, the difference shows the same ``sign.''



Finally, for the eight Ni lines in common with \gh, we plot our normalized spectra 
along with spectrum syntheses (see Figures~\ref{fig:ni_syntha} and \ref{fig:ni_synthb}). The main points to take from 
these figures are the location of the continuum and how well the 
spectrum syntheses fit the lines for the abundances we 
measure. We note that our abundances were determined from 
EW analysis rather than spectrum synthesis. Nevertheless, had we relied solely upon 
spectrum synthesis, we would have obtained essentially identical results. 
A systematic increase in $\log\,\epsilon$(Ni) of 0.17\,dex or in [Ni/Fe] of 0.09 dex, 
as measured by GH09, is not supported by these spectrum syntheses. 

The main conclusions we draw from this comparison are 
(i) abundance differences between the two studies cannot be attributed 
to the different versions of model atmospheres and spectrum synthesis software; 
(ii) the [Ni/Fe] ratio remains unchanged when using our line list but with 
either the GH09 stellar parameters or our stellar parameters; 
(iii) differences in [Ni/Fe] cannot be attributed to the adopted solar abundances;
(iv) although the set of lines in common between the two analyses is small, 
there are no large systematic differences in the $\log\,{gf}$ values that could 
explain the discrepancy in Ni abundances;
(v) for \fei\ and \feii, our EWs are systematically lower than those of GH09 by $\sim 6$\,m\AA, and our Ni EWs are systematically lower by $\sim 12$\,m\AA; and
(vi) our EW uncertainties for Ni are consistent with the observed dispersion in Ni abundance.

As noted above, while our measured [Ni/Fe] value does not represent a substantial revision of the 
GH09 value, our Ni abundance is not unusual with respect to field stars 
at the same metallicity.
Nevertheless, we welcome further analyses of this star,  
preferably conducted with higher-quality spectra.

\begin{deluxetable}{lccc}
\tabletypesize{\scriptsize}
\tablecolumns{4} 
\tablewidth{0pc} 
\tablecaption{Comparison of Ni measurement \label{tab:ni_comparison}}
\tablehead{ 
\colhead{Species} & 			
\colhead{$\log \epsilon (X)$} & 	
\colhead{[X/H]\tablenotemark{a}} & 	
\colhead{$\sigma$} \\			
\colhead{(1)} &
\colhead{(2)} &
\colhead{(3)} 
}
\startdata 
 \noalign{\vskip +0.5ex}
 \multicolumn{4}{c}{This work} \cr
 \noalign{\vskip  .8ex}%
 \hline
 \noalign{\vskip-2ex}\\
Fe I  & 7.37 & -0.13 & 0.18 \\
Fe II  & 7.35  & -0.15 & 0.18 \\
Ni I & 6.16   & -0.06 & 0.14  \\
\cutinhead{Measurements from \gh}
Fe I  & 7.42  & -0.08 & 0.10 \\
Fe II  & 7.42 & -0.08 & 0.12 \\
Ni I  & 6.36 & 0.16 & 0.19  \\
\cutinhead{This work using \gh's line list and stellar parameters}
Fe I  & 7.38   & -0.12 & 0.13 \\
Fe II  & 7.42 & -0.08 & 0.10 \\
Ni I  & 6.33 & 0.11 & 0.19  \\
\cutinhead{This work using \gh's line list and this work's stellar parameters}
Fe I  & 7.42   & -0.08 & 0.12 \\
Fe II  & 7.42 & -0.08 & 0.10 \\
Ni I  & 6.36 & 0.14 & 0.19  \\

\tablenotetext{a}{Using  solar values from \citet{2009ARA&A..47..481A}.}

\enddata


\end{deluxetable}

\begin{deluxetable}{cccrrrrrr}
\tabletypesize{\scriptsize}
\tablecolumns{9} 
\tablewidth{0pc} 
\tablecaption{Line List and Equivalent Width Measurements \label{tab:ew}}
\tablehead{ 
\colhead{Wavelength (\AA)} &          
\colhead{Species} &   		      
\colhead{$\chi$ (eV)} &               
\colhead{$\log gf$} &                 
\colhead{Star A} &                    
\colhead{Star C} &                    
\colhead{Star E} &                    
\colhead{Star G} &                    
\colhead{Source} \\                   
\cline{5-8} 
\colhead{} &          		
\colhead{} &   		      	
\colhead{} &               	
\colhead{} &                 	
\multicolumn{4}{c}{EW (m\AA)} \\
\colhead{(1)} &
\colhead{(2)} &
\colhead{(3)} &
\colhead{(4)} &
\colhead{(5)} &
\colhead{(6)} &
\colhead{(7)} &
\colhead{(8)} & 
\colhead{(9)} 
}
\startdata 
  4602.00 & 26.0 &  1.607 & $-$3.154 &  114.8 & \ldots & \ldots & \ldots &    Oxford \\
  4733.59 & 26.0 &  1.484 & $-$2.987 & \ldots & \ldots &  102.3 &   85.4 &    Oxford \\
  4802.88 & 26.0 &  3.692 & $-$1.531 &   91.0 &   96.0 & \ldots &   64.8 &    Oxford \\
  4848.88 & 26.0 &  2.277 & $-$3.154 & \ldots &   85.9 & \ldots & \ldots &    Oxford \\
  4930.31 & 26.0 &  3.957 & $-$1.264 & \ldots & \ldots & \ldots &   57.8 &    Oxford \\
  4962.57 & 26.0 &  4.175 & $-$1.199 &   77.5 & \ldots & \ldots & \ldots &    Oxford \\
  5014.94 & 26.0 &  3.940 & $-$0.320 & \ldots & \ldots & \ldots &  106.8 &    Oxford \\
  5044.21 & 26.0 &  2.849 & $-$2.034 &  109.1 & \ldots & \ldots &   67.2 &    Oxford \\
  5049.82 & 26.0 &  2.277 & $-$1.372 & \ldots & \ldots & \ldots &  117.8 &    Oxford \\
  5054.64 & 26.0 &  3.637 & $-$1.938 &   70.8 & \ldots & \ldots &   30.4 &    Oxford \\
  5083.34 & 26.0 &  0.957 & $-$2.958 & \ldots & \ldots & \ldots &   96.2 &    Oxford \\
  5141.74 & 26.0 &  2.422 & $-$2.001 &  123.9 & \ldots & \ldots &   85.4 &    Oxford \\
  5151.91 & 26.0 &  1.010 & $-$3.322 & \ldots & \ldots & \ldots &   85.0 &    Oxford \\
  5166.28 & 26.0 &  0.000 & $-$4.195 & \ldots & \ldots &  111.6 & \ldots &    Oxford \\
  5194.94 & 26.0 &  1.556 & $-$2.090 & \ldots & \ldots & \ldots &  113.8 &    Oxford \\
  5198.71 & 26.0 &  2.221 & $-$2.135 & \ldots & \ldots & \ldots &   92.1 &    Oxford \\
  5217.39 & 26.0 &  3.209 & $-$1.179 & \ldots & \ldots &  123.7 &   81.9 &    Oxford \\
  5223.19 & 26.0 &  3.632 & $-$1.800 &   58.5 & \ldots &   33.9 &   27.7 &    Oxford \\
  5225.52 & 26.0 &  0.110 & $-$4.789 & \ldots & \ldots &   82.3 &   65.6 &    Oxford \\
  5242.49 & 26.0 &  3.632 & $-$0.984 &  113.9 & \ldots &  110.6 &   69.8 &    Oxford \\
  5247.05 & 26.0 &  0.087 & $-$4.946 & \ldots & \ldots &   76.6 &   55.5 &    Oxford \\
  5250.21 & 26.0 &  0.121 & $-$4.938 & \ldots & \ldots & \ldots &   59.9 &    Oxford \\
  5253.46 & 26.0 &  3.281 & $-$1.630 & \ldots & \ldots & \ldots &   62.4 &    Oxford \\
  5412.80 & 26.0 &  4.431 & $-$1.783 &   38.2 & \ldots & \ldots & \ldots &    Oxford \\
  5491.84 & 26.0 &  4.183 & $-$2.253 &   36.9 & \ldots & \ldots & \ldots &    Oxford \\
  5497.52 & 26.0 &  1.010 & $-$2.849 & \ldots & \ldots &  129.3 &  102.1 &    Oxford \\
  5501.46 & 26.0 &  0.957 & $-$3.063 & \ldots & \ldots & \ldots &   90.9 &    Oxford \\
  5506.78 & 26.0 &  0.989 & $-$2.797 & \ldots & \ldots & \ldots &   95.7 &    Oxford \\
  5525.54 & 26.0 &  4.227 & $-$1.149 & \ldots & \ldots & \ldots &   38.4 &    Oxford \\
  5569.62 & 26.0 &  3.414 & $-$0.544 & \ldots & \ldots &  123.1 &  113.4 &    Oxford \\
  5586.76 & 26.0 &  3.366 & $-$0.161 & \ldots & \ldots &  142.8 & \ldots &    Oxford \\
  5600.23 & 26.0 &  4.257 & $-$1.486 & \ldots & \ldots & \ldots &   44.2 &    Oxford \\
  5618.63 & 26.0 &  4.206 & $-$1.292 &   80.1 & \ldots & \ldots &   43.3 &    Oxford \\
  5661.35 & 26.0 &  4.281 & $-$1.822 &   48.2 & \ldots & \ldots &   18.0 &    Oxford \\
  5662.51 & 26.0 &  4.175 & $-$0.590 &  115.6 & \ldots &   79.1 &   69.5 &    Oxford \\
  5701.55 & 26.0 &  2.557 & $-$2.216 & \ldots & \ldots &  104.7 & \ldots &    Oxford \\
  5705.47 & 26.0 &  4.298 & $-$1.421 &   78.9 &   69.5 & \ldots & \ldots &    Oxford \\
  5741.85 & 26.0 &  4.253 & $-$1.689 &   58.0 & \ldots & \ldots & \ldots &    Oxford \\
  5753.12 & 26.0 &  4.257 & $-$0.705 & \ldots & \ldots & \ldots &   52.0 &    Oxford \\
  5775.08 & 26.0 &  4.217 & $-$1.314 &   82.3 &   73.0 & \ldots &   34.5 &    Oxford \\
  5778.45 & 26.0 &  2.586 & $-$3.481 &   55.5 &   58.5 & \ldots & \ldots &    Oxford \\
  5816.37 & 26.0 &  4.545 & $-$0.618 & \ldots & \ldots & \ldots &   66.5 &    Oxford \\
  5855.09 & 26.0 &  4.604 & $-$1.547 &   43.1 & \ldots &   17.1 & \ldots &    Oxford \\
  5909.97 & 26.0 &  3.209 & $-$2.643 & \ldots &   44.9 & \ldots &   36.5 &    Oxford \\
  5916.25 & 26.0 &  2.452 & $-$2.994 & \ldots &   73.6 & \ldots &   44.1 &    Oxford \\
  5956.69 & 26.0 &  0.858 & $-$4.608 &  107.2 &  101.5 &   59.7 &   34.9 &    Oxford \\
  6012.21 & 26.0 &  2.221 & $-$4.073 &   72.0 & \ldots & \ldots & \ldots &    Oxford \\
  6027.05 & 26.0 &  4.073 & $-$1.106 &  100.3 &   79.9 & \ldots &   63.3 &    Oxford \\
  6065.48 & 26.0 &  2.607 & $-$1.530 & \ldots & \ldots &  106.2 & \ldots &    Oxford \\
  6082.71 & 26.0 &  2.221 & $-$3.573 & \ldots &   61.6 & \ldots & \ldots &    Oxford \\
  6120.24 & 26.0 &  0.914 & $-$5.970 &   33.9 & \ldots & \ldots & \ldots &    Oxford \\
  6136.62 & 26.0 &  2.452 & $-$1.400 & \ldots & \ldots &  145.6 & \ldots &    Oxford \\
  6136.99 & 26.0 &  2.196 & $-$2.950 & \ldots & \ldots &   71.7 & \ldots &    Oxford \\
  6151.62 & 26.0 &  2.174 & $-$3.299 &   91.4 &   81.8 &   33.7 &   36.6 &    Oxford \\
  6165.36 & 26.0 &  4.140 & $-$1.490 &   73.5 & \ldots &   54.0 & \ldots &    Oxford \\
  6173.34 & 26.0 &  2.221 & $-$2.880 &  115.1 & \ldots & \ldots &   52.8 &    Oxford \\
  6180.20 & 26.0 &  2.725 & $-$2.637 & \ldots &   61.0 & \ldots &   46.8 &    Oxford \\
  6200.31 & 26.0 &  2.607 & $-$2.437 & \ldots &  109.6 & \ldots &   66.6 &    Oxford \\
  6219.28 & 26.0 &  2.196 & $-$2.433 & \ldots &  134.4 & \ldots &   75.2 &    Oxford \\
  6229.23 & 26.0 &  2.843 & $-$2.846 &   82.3 & \ldots &   61.5 & \ldots &    Oxford \\
  6230.73 & 26.0 &  2.557 & $-$1.281 & \ldots & \ldots & \ldots &  109.9 &    Oxford \\
  6232.64 & 26.0 &  3.651 & $-$1.283 &  114.5 &  129.1 &  106.6 &   65.3 &    Oxford \\
  6246.32 & 26.0 &  3.600 & $-$0.894 & \ldots & \ldots & \ldots &  106.2 &    Oxford \\
  6252.55 & 26.0 &  2.402 & $-$1.687 & \ldots & \ldots & \ldots &  100.8 &    Oxford \\
  6265.13 & 26.0 &  2.174 & $-$2.550 & \ldots & \ldots & \ldots &   75.0 &    Oxford \\
  6270.22 & 26.0 &  2.856 & $-$2.505 & \ldots &   77.1 &   70.2 & \ldots &    Oxford \\
  6297.79 & 26.0 &  2.221 & $-$2.740 &  120.6 &  115.1 &   93.5 &   61.7 &    Oxford \\
  6301.50 & 26.0 &  3.651 & $-$0.766 & \ldots & \ldots &  114.0 &  110.9 &    Oxford \\
  6322.69 & 26.0 &  2.586 & $-$2.426 &  120.9 &  119.1 & \ldots & \ldots &    Oxford \\
  6335.33 & 26.0 &  2.196 & $-$2.194 & \ldots & \ldots &  116.4 &   75.0 &    Oxford \\
  6336.82 & 26.0 &  3.684 & $-$0.916 & \ldots & \ldots & \ldots &   69.1 &    Oxford \\
  6344.15 & 26.0 &  2.431 & $-$2.923 & \ldots & \ldots & \ldots &   48.4 &    Oxford \\
  6355.03 & 26.0 &  2.843 & $-$2.403 & \ldots &  101.7 &   93.4 &   64.4 &    Oxford \\
  6393.60 & 26.0 &  2.431 & $-$1.469 & \ldots & \ldots &  129.4 &  105.5 &    Oxford \\
  6408.02 & 26.0 &  3.684 & $-$1.066 & \ldots &  127.1 &   87.4 &   80.7 &    Oxford \\
  6411.65 & 26.0 &  3.651 & $-$0.734 & \ldots & \ldots & \ldots &  118.0 &    Oxford \\
  6430.84 & 26.0 &  2.174 & $-$2.006 & \ldots & \ldots &  135.0 &  102.5 &    Oxford \\
  6481.87 & 26.0 &  2.277 & $-$2.984 &  113.2 & \ldots & \ldots &   50.3 &    Oxford \\
  6498.94 & 26.0 &  0.957 & $-$4.687 & \ldots &  121.1 & \ldots &   35.3 &    Oxford \\
  6574.23 & 26.0 &  0.989 & $-$5.004 &   84.3 &   88.5 & \ldots &   25.3 &    Oxford \\
  6575.02 & 26.0 &  2.586 & $-$2.727 &  108.2 &  115.3 & \ldots &   51.0 &    Oxford \\
  6592.91 & 26.0 &  2.725 & $-$1.490 & \ldots & \ldots &  121.7 &  104.1 &    Oxford \\
  6593.87 & 26.0 &  2.431 & $-$2.422 & \ldots & \ldots &   99.1 &   75.4 &    Oxford \\
  6609.11 & 26.0 &  2.557 & $-$2.692 & \ldots &  104.9 &   54.7 &   53.1 &    Oxford \\
  6648.08 & 26.0 &  1.010 & $-$5.918 &   48.2 & \ldots & \ldots & \ldots &    Oxford \\
  6677.99 & 26.0 &  2.690 & $-$1.435 & \ldots & \ldots &  142.4 &   98.3 &    Oxford \\
  6699.16 & 26.0 &  4.590 & $-$2.170 &   24.6 &   15.8 & \ldots & \ldots &    Oxford \\
  6739.52 & 26.0 &  1.556 & $-$4.823 &   48.4 &   53.3 & \ldots & \ldots &    Oxford \\
  6750.15 & 26.0 &  2.422 & $-$2.621 &  120.0 &   96.6 &   70.2 &   60.0 &    Oxford \\
  6752.70 & 26.0 &  4.635 & $-$1.273 & \ldots & \ldots & \ldots &   33.8 &    Oxford \\
  6810.26 & 26.0 &  4.603 & $-$1.003 &   73.7 & \ldots & \ldots &   37.5 &    Oxford \\
  6837.02 & 26.0 &  4.590 & $-$1.756 & \ldots &   19.0 &   21.8 & \ldots &    Oxford \\
  7112.17 & 26.0 &  2.988 & $-$3.044 &   86.4 &   48.9 &   40.7 & \ldots &    Oxford \\
  7223.66 & 26.0 &  3.015 & $-$2.269 & \ldots &   87.5 &   49.9 &   35.0 &    Oxford \\
  7401.69 & 26.0 &  4.183 & $-$1.664 & \ldots &   38.8 &   30.4 & \ldots &    Oxford \\
  7710.36 & 26.0 &  4.217 & $-$1.129 & \ldots &   66.5 &   56.2 & \ldots &    Oxford \\
  7723.20 & 26.0 &  2.277 & $-$3.617 & \ldots &   86.2 &   37.3 & \ldots &    Oxford \\
  7912.86 & 26.0 &  0.858 & $-$4.848 & \ldots &   97.1 & \ldots &   21.7 &    Oxford \\
  7941.09 & 26.0 &  3.271 & $-$2.331 &   76.6 &   42.5 & \ldots &   23.9 &    Oxford \\
  8075.15 & 26.0 &  0.914 & $-$5.088 &  105.4 &  118.7 & \ldots & \ldots &    Oxford \\
  4491.40 & 26.1 &  2.853 & $-$2.684 & \ldots & \ldots &  106.6 & \ldots & Biemont93 \\
  4508.29 & 26.1 &  2.853 & $-$2.312 &  104.1 &  106.2 & \ldots &  108.0 & Biemont93 \\
  4620.52 & 26.1 &  2.826 & $-$3.079 &   75.7 & \ldots & \ldots &   69.7 & Biemont93 \\
  4635.32 & 26.1 &  5.952 & $-$1.275 & \ldots & \ldots &   48.1 &   23.9 & Biemont93 \\
  4993.36 & 26.1 &  2.805 & $-$3.485 &   58.4 & \ldots & \ldots & \ldots & Biemont93 \\
  5100.66 & 26.1 &  2.805 & $-$4.135 & \ldots & \ldots & \ldots &   42.1 & Biemont93 \\
  5132.67 & 26.1 &  2.805 & $-$3.901 &   41.5 & \ldots & \ldots & \ldots & Biemont93 \\
  5197.58 & 26.1 &  3.228 & $-$2.233 &  101.2 & \ldots & \ldots &   86.8 & Biemont93 \\
  5234.62 & 26.1 &  3.219 & $-$2.151 &  105.4 & \ldots & \ldots &   93.4 & Biemont93 \\
  5414.07 & 26.1 &  3.219 & $-$3.750 &   45.8 & \ldots & \ldots & \ldots & Biemont93 \\
  5425.26 & 26.1 &  3.197 & $-$3.372 &   58.1 &   63.6 & \ldots & \ldots & Biemont93 \\
  5991.38 & 26.1 &  3.150 & $-$3.557 & \ldots &   30.0 &   54.4 &   23.5 & Biemont93 \\
  6084.11 & 26.1 &  3.197 & $-$3.808 &   37.3 & \ldots & \ldots & \ldots & Biemont93 \\
  6149.26 & 26.1 &  3.886 & $-$2.724 &   48.1 & \ldots &   53.2 &   38.2 & Biemont93 \\
  6239.95 & 26.1 &  3.886 & $-$3.439 & \ldots & \ldots & \ldots &   25.8 & Biemont93 \\
  6247.56 & 26.1 &  3.889 & $-$2.329 &   60.4 & \ldots &   74.7 &   68.7 & Biemont93 \\
  6369.46 & 26.1 &  2.889 & $-$4.253 &   38.5 &   28.2 &   41.7 &   24.1 & Biemont93 \\
  6383.72 & 26.1 &  5.548 & $-$2.271 & \ldots & \ldots & \ldots &   12.4 & Biemont93 \\
  6432.68 & 26.1 &  2.889 & $-$3.708 &   62.8 &   39.0 & \ldots & \ldots & Biemont93 \\
  6456.38 & 26.1 &  3.900 & $-$2.075 & \ldots & \ldots & \ldots &   75.7 & Biemont93 \\
  6516.08 & 26.1 &  2.889 & $-$3.450 & \ldots & \ldots & \ldots &   54.7 & Biemont93 \\
  7222.39 & 26.1 &  3.886 & $-$3.295 & \ldots & \ldots & \ldots &   20.0 & Biemont93 \\
  7711.72 & 26.1 &  3.900 & $-$2.543 &   68.6 & \ldots &   85.7 &   58.9 & Biemont93 \\
  5082.35 & 28.0 &  3.660 & $-$0.590 &   87.3 & \ldots &   69.2 &   65.5 &   Reddy03 \\
  5088.54 & 28.0 &  3.850 & $-$1.040 &   53.2 & \ldots & \ldots &   30.2 &   Reddy03 \\
  5088.96 & 28.0 &  3.680 & $-$1.240 &   53.4 & \ldots & \ldots & \ldots &   Reddy03 \\
  5094.42 & 28.0 &  3.830 & $-$1.070 &   48.3 & \ldots & \ldots & \ldots &   Reddy03 \\
  5115.40 & 28.0 &  3.830 & $-$0.280 &   89.1 & \ldots & \ldots & \ldots &   Reddy03 \\
  5682.20 & 28.0 &  4.100 & $-$0.470 &   75.5 & \ldots & \ldots & \ldots &      RC02 \\
  5748.35 & 28.0 &  1.680 & $-$3.260 &   74.7 &   62.2 & \ldots & \ldots &      RC02 \\
  5749.30 & 28.0 &  3.940 & $-$1.990 &   17.1 & \ldots & \ldots & \ldots &      RC02 \\
  5847.01 & 28.0 &  1.680 & $-$3.410 &   64.9 &   39.5 & \ldots & \ldots &   Reddy03 \\
  6007.31 & 28.0 &  1.680 & $-$3.340 &   66.0 &   37.9 & \ldots & \ldots &      RC02 \\
  6053.69 & 28.0 &  4.230 & $-$1.070 &   40.5 & \ldots & \ldots & \ldots &      RC02 \\
  6086.28 & 28.0 &  4.260 & $-$0.520 &   61.9 & \ldots & \ldots &   34.0 &      RC02 \\
  6108.12 & 28.0 &  1.680 & $-$2.440 &  111.8 &   94.6 & \ldots &   53.6 &      RC02 \\
  6111.08 & 28.0 &  4.090 & $-$0.810 &   60.9 &   22.5 & \ldots & \ldots &   Reddy03 \\
  6130.14 & 28.0 &  4.270 & $-$0.940 &   43.9 & \ldots & \ldots & \ldots &   Reddy03 \\
  6175.37 & 28.0 &  4.090 & $-$0.550 &   69.6 &   55.7 & \ldots &   46.8 &   Reddy03 \\
  6176.82 & 28.0 &  4.090 & $-$0.260 &   85.8 &   57.4 &   56.7 &   57.4 &   Reddy03 \\
  6177.25 & 28.0 &  1.830 & $-$3.510 &   50.3 & \ldots & \ldots & \ldots &   Reddy03 \\
  6186.71 & 28.0 &  4.100 & $-$0.970 &   52.7 & \ldots & \ldots &   21.6 &      RC02 \\
  6204.61 & 28.0 &  4.090 & $-$1.110 &   46.2 & \ldots & \ldots & \ldots &   Reddy03 \\
  6322.17 & 28.0 &  4.150 & $-$1.170 &   39.5 & \ldots & \ldots & \ldots &      RC02 \\
  6370.35 & 28.0 &  3.540 & $-$1.940 &   38.4 & \ldots & \ldots & \ldots &      RC02 \\
  6378.26 & 28.0 &  4.150 & $-$0.830 &   56.1 &   40.9 & \ldots & \ldots &   Reddy03 \\
  6482.80 & 28.0 &  1.930 & $-$2.630 &   88.8 &   64.4 & \ldots &   24.1 &      RC02 \\
  6598.60 & 28.0 &  4.230 & $-$0.980 &   44.8 &   24.9 & \ldots & \ldots &      RC02 \\
  6635.12 & 28.0 &  4.420 & $-$0.830 &   45.1 & \ldots & \ldots & \ldots &      RC02 \\
  6643.64 & 28.0 &  1.680 & $-$2.030 &  151.0 &  127.3 &   95.1 &   71.8 &   Reddy03 \\
  6767.77 & 28.0 &  1.830 & $-$2.170 &  127.1 &  129.1 & \ldots &   66.9 &      RC02 \\
  6772.32 & 28.0 &  3.660 & $-$0.970 &   78.0 & \ldots & \ldots & \ldots &   Reddy03 \\
  6842.04 & 28.0 &  3.660 & $-$1.470 &   58.1 & \ldots & \ldots &   23.9 &      RC02 \\
  7030.01 & 28.0 &  3.540 & $-$1.730 &   53.6 & \ldots & \ldots & \ldots &      RC02 \\
  7122.20 & 28.0 &  3.540 &    0.050 &  146.4 &  120.8 & \ldots & \ldots &      RC02 \\
  7261.92 & 28.0 &  1.950 & $-$2.700 & \ldots & \ldots & \ldots &   38.1 &      RC02 \\
  7327.65 & 28.0 &  3.800 & $-$1.770 &   35.1 & \ldots & \ldots &   13.3 &      RC02 \\
  7409.35 & 28.0 &  3.800 & $-$0.100 & \ldots &   83.6 & \ldots & \ldots &      RC02 \\
  7414.50 & 28.0 &  1.990 & $-$2.570 & \ldots &   83.6 &   48.0 &   40.4 &      RC02 \\
  7422.28 & 28.0 &  3.630 & $-$0.130 &  139.5 &  101.4 &   87.4 & \ldots &      RC02 \\
  7574.05 & 28.0 &  3.830 & $-$0.580 &   91.2 &   39.5 & \ldots &   41.8 &      RC02 \\
  7748.89 & 28.0 &  3.700 & $-$0.380 &  117.7 &   85.7 &   95.7 &   70.3 &   Reddy03 \\
  7788.93 & 28.0 &  1.950 & $-$2.420 & \ldots &  123.4 &   82.8 & \ldots &      RC02 \\
  7797.59 & 28.0 &  3.900 & $-$0.350 &  106.4 &   84.0 &   81.1 &   61.7 &   Reddy03 \\
  7917.44 & 28.0 &  3.740 & $-$1.500 & \ldots & \ldots & \ldots &   14.8 &      RC02 \\
\enddata

\tablecomments{Oxford = 
\citet{1979MNRAS.186..657B,1979MNRAS.186..633B,1980MNRAS.191..445B,1986MNRAS.220..549B,1995A&A...296..217B}}.

\end{deluxetable}

\begin{deluxetable}{crrrr}
\tabletypesize{\scriptsize}
\tablecolumns{5} 
\tablewidth{0pc} 
\tablecaption{Equivalent Width Uncertainties for Ni in Star G\label{tab:err_g}}
\tablehead{ 
\colhead{Wavelength (\AA)} &				
\colhead{EW} &						
\colhead{$\sigma_1$\tablenotemark{a}} &			
\colhead{$\sigma_2$\tablenotemark{b}} &			
\colhead{$\sigma_{\rm Total}$\tablenotemark{c}}  \\ 	
\colhead{(1)} &
\colhead{(2)} &
\colhead{(3)} &
\colhead{(4)} &
\colhead{(5)} 
}
\startdata 
5082.35 & 55.1 & 5.0 & 1.2 & 5.1 \\
5088.54 & 25.7 & 5.0 & 0.8 & 5.0 \\
6086.28 & 33.5 & 6.0 & 1.3 & 6.1 \\
6108.12 & 58.1 & 5.8 & 1.1 & 5.9 \\
6175.37 & 39.8 & 6.4 & 1.1 & 6.5 \\
6176.82 & 54.4 & 6.0 & 1.0 & 6.0 \\
6186.71 & 21.1 & 5.2 & 1.9 & 5.6 \\
6482.80 & 37.6 & 6.0 & 0.7 & 6.0 \\
6643.64 & 79.9 & 6.2 & 1.1 & 6.3 \\
6767.77 & 66.9 & 6.8 & 2.4 & 7.2 \\
6842.04 & 18.7 & 6.3 & 1.7 & 6.6 \\
7261.92 & 35.6 & 6.7 & 1.3 & 6.9 \\
7327.65 &  8.4 & 6.4 & 1.5 & 6.6 \\
7414.50 & 40.9 & 7.0 & 0.2 & 7.0 \\
7574.05 & 53.8 & 6.4 & 0.8 & 6.5 \\
7748.89 & 71.3 & 7.4 & 0.7 & 7.5 \\
7797.59 & 63.3 & 7.7 & 1.4 & 7.8 \\
7917.44 & 16.8 & 7.5 & 0.4 & 7.5 \\
\enddata

\tablenotetext{a}{This is the error from the \citet{1988IAUS..132..345C} formula.}
\tablenotetext{b}{This is the error due to continuum placement 
(see text for details).}
\tablenotetext{c}{This is the total error obtained by 
adding columns (4) and (5) in quadrature.}

\end{deluxetable}

\begin{figure*}[t!]
\epsscale{0.8}
\vspace{5mm}
\includegraphics[width=\textwidth]{\plotdir 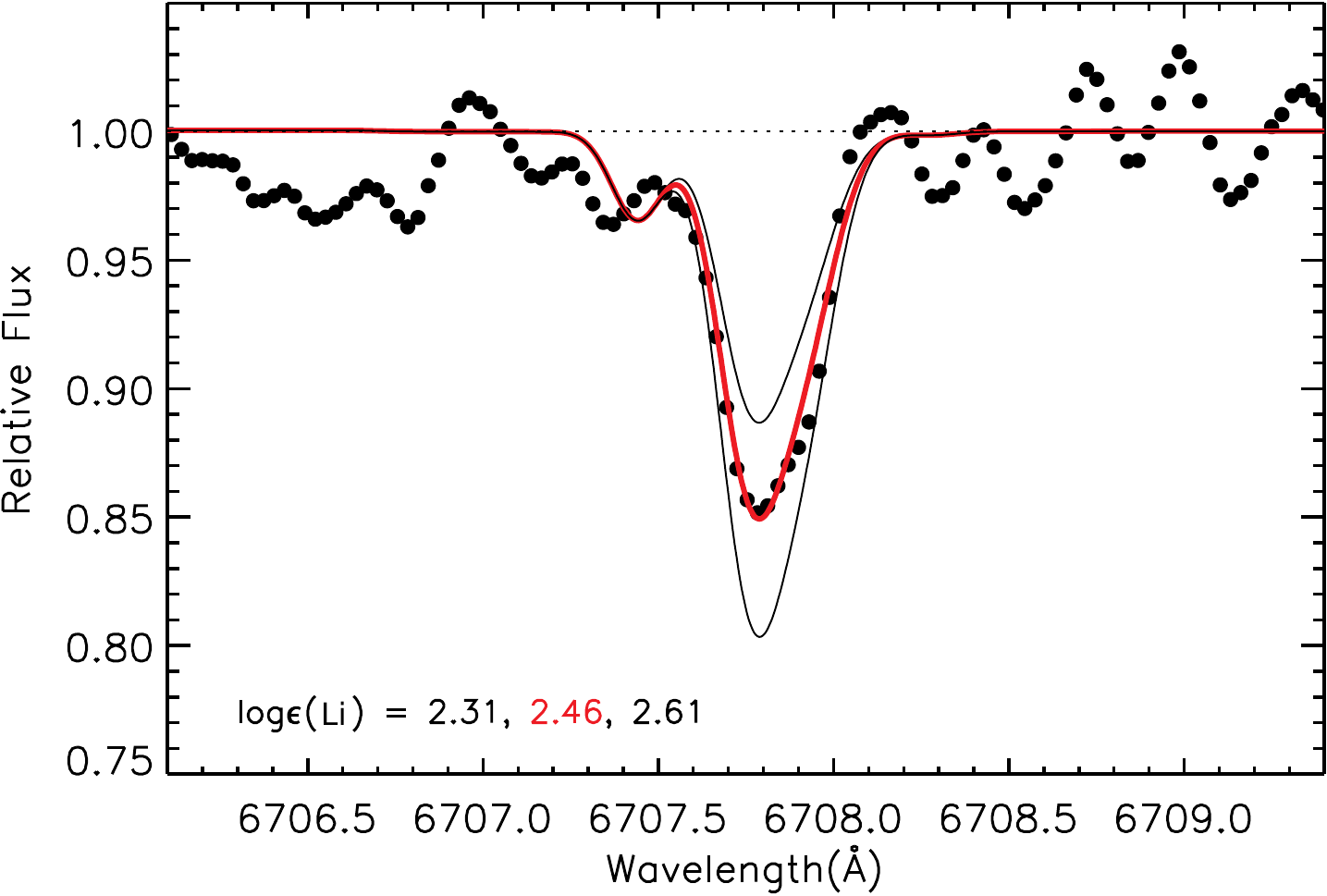} 
\caption{Observed spectra of Tycho-G centered around the Li $\lambda$6707 line. 
Synthetic spectra with different Li abundances are overplotted. 
The thick red line represents the Li abundance corresponding to the 
best-fitting value, and unsatisfactory fits ($\pm 0.15$\,dex) are 
plotted as thin black lines. 
\label{fig:li_synth}}
\end{figure*}

\begin{figure*}[t!]
\epsscale{0.8}
\vspace{5mm}
\includegraphics[width=\textwidth]{\plotdir 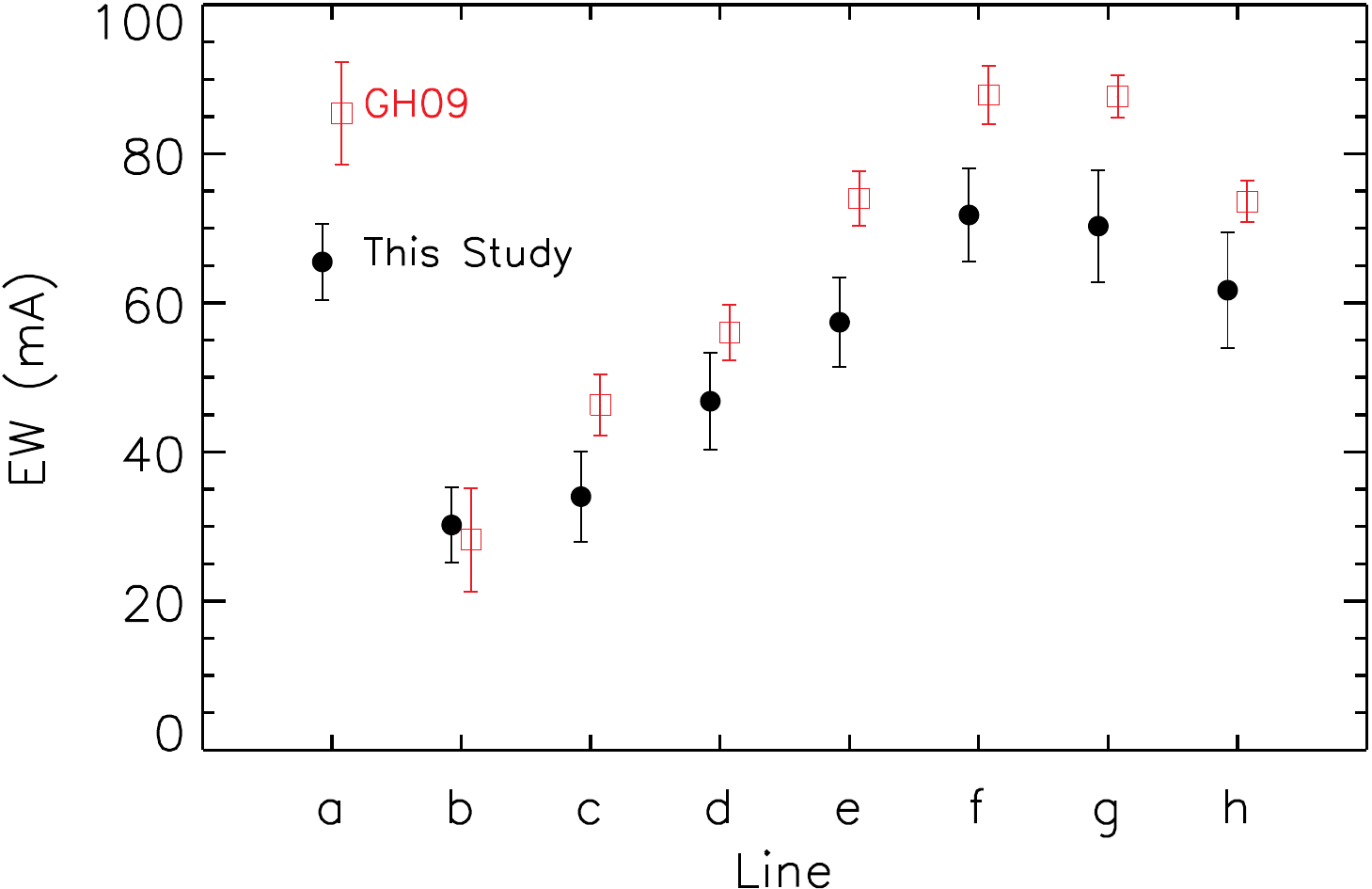} 
\caption{EWs for the eight Ni lines in common between GH09 (open red squares) 
and this study (filled black circles) for Tycho-G. Lines (a-h) 
are 
5082.35\,\AA, 
5088.54\,\AA, 
6086.28\,\AA, 
6175.37\,\AA, 
6176.82\,\AA, 
6643.64\,\AA, 
7748.89\,\AA, and  
7797.59\,\AA, respectively. 
\label{fig:ew_comp}}
\end{figure*}

\begin{figure*}[ht!]
\epsscale{0.9}
\vspace{5mm}
\includegraphics[width=\textwidth]{\plotdir 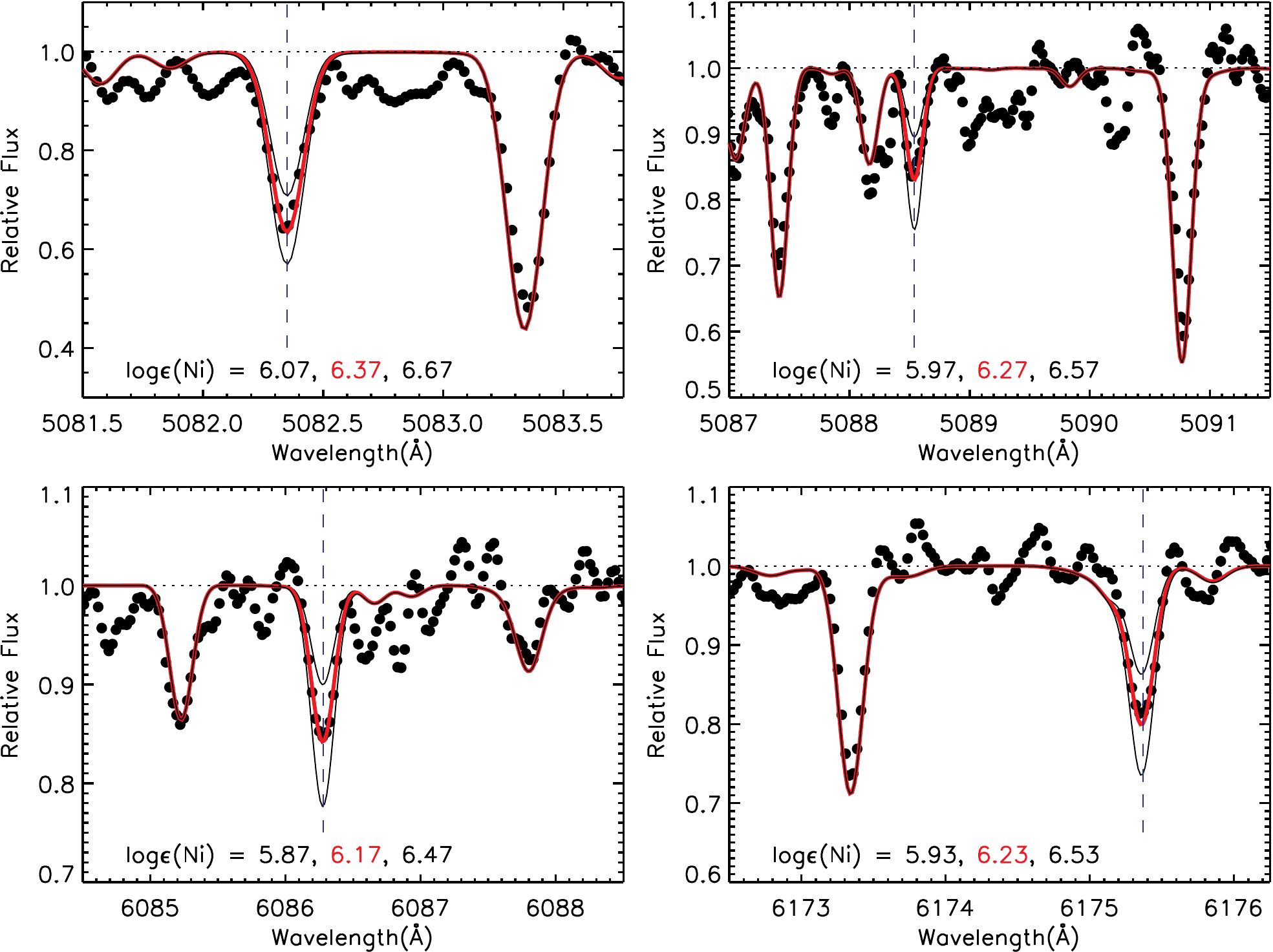} 
\caption{Observed spectra centered around five Ni lines in common with GH09
for Tycho-G. 
Synthetic spectra with different Ni abundances are overplotted. 
The thick red line represents the Ni abundance corresponding to the value 
derived from EW analysis, and unsatisfactory fits ($\pm 0.3$ dex) are 
plotted as thin black lines. 
\label{fig:ni_syntha}}
\end{figure*}

\begin{figure*}[ht!]
\epsscale{0.9}
\vspace{5mm}
\includegraphics[width=\textwidth]{\plotdir 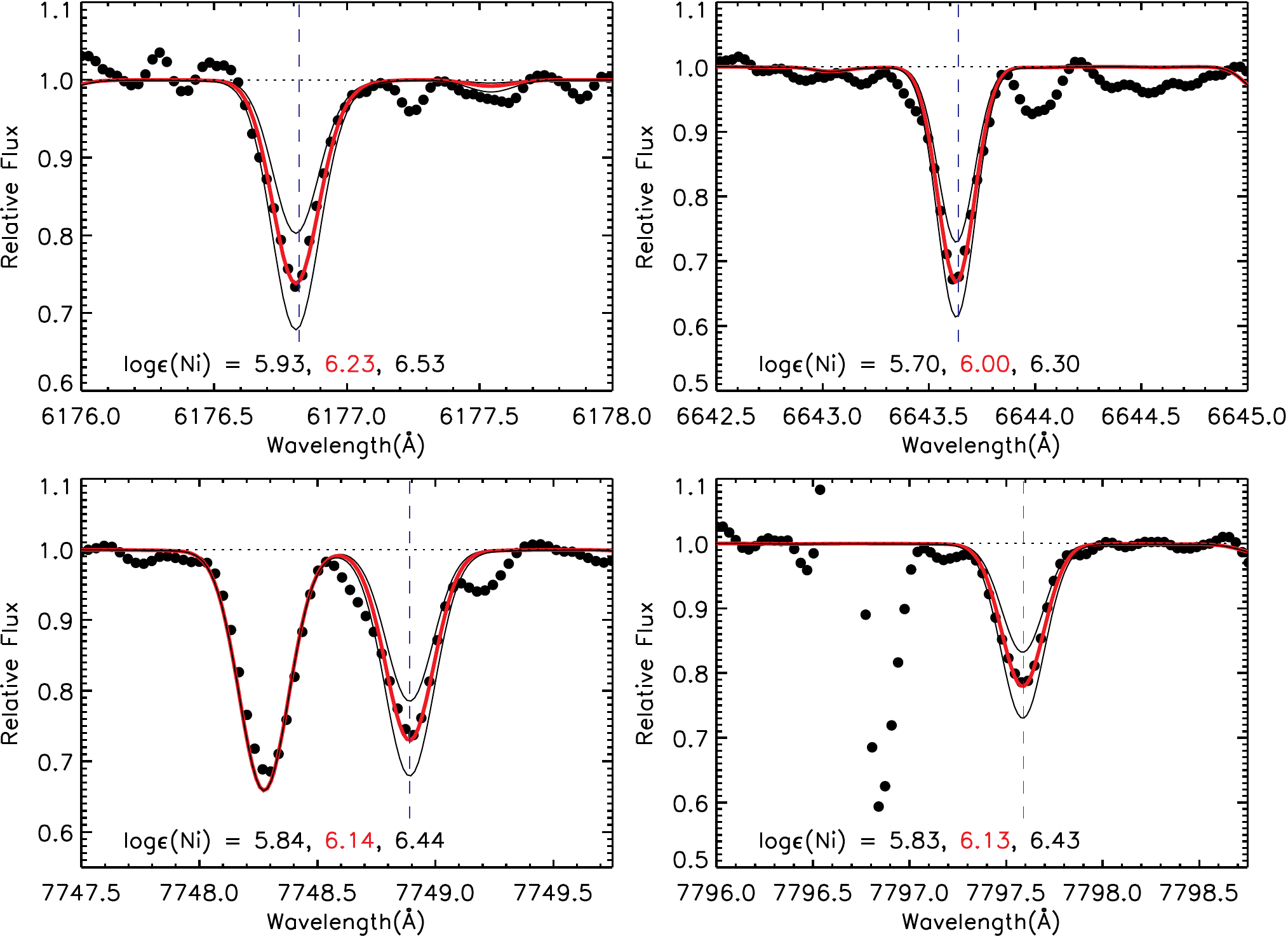} 
\caption{Same as Figure \ref{fig:ni_syntha} but for 
the remaining four Ni lines in common with GH09 (the upper-left line is also seen in the previous panel). 
\label{fig:ni_synthb}}
\end{figure*}

\subsection{Distances}
\label{sec:distance}
\begin{figure*}[ht!] 
   \includegraphics[width=0.5\textwidth]{\plotdir 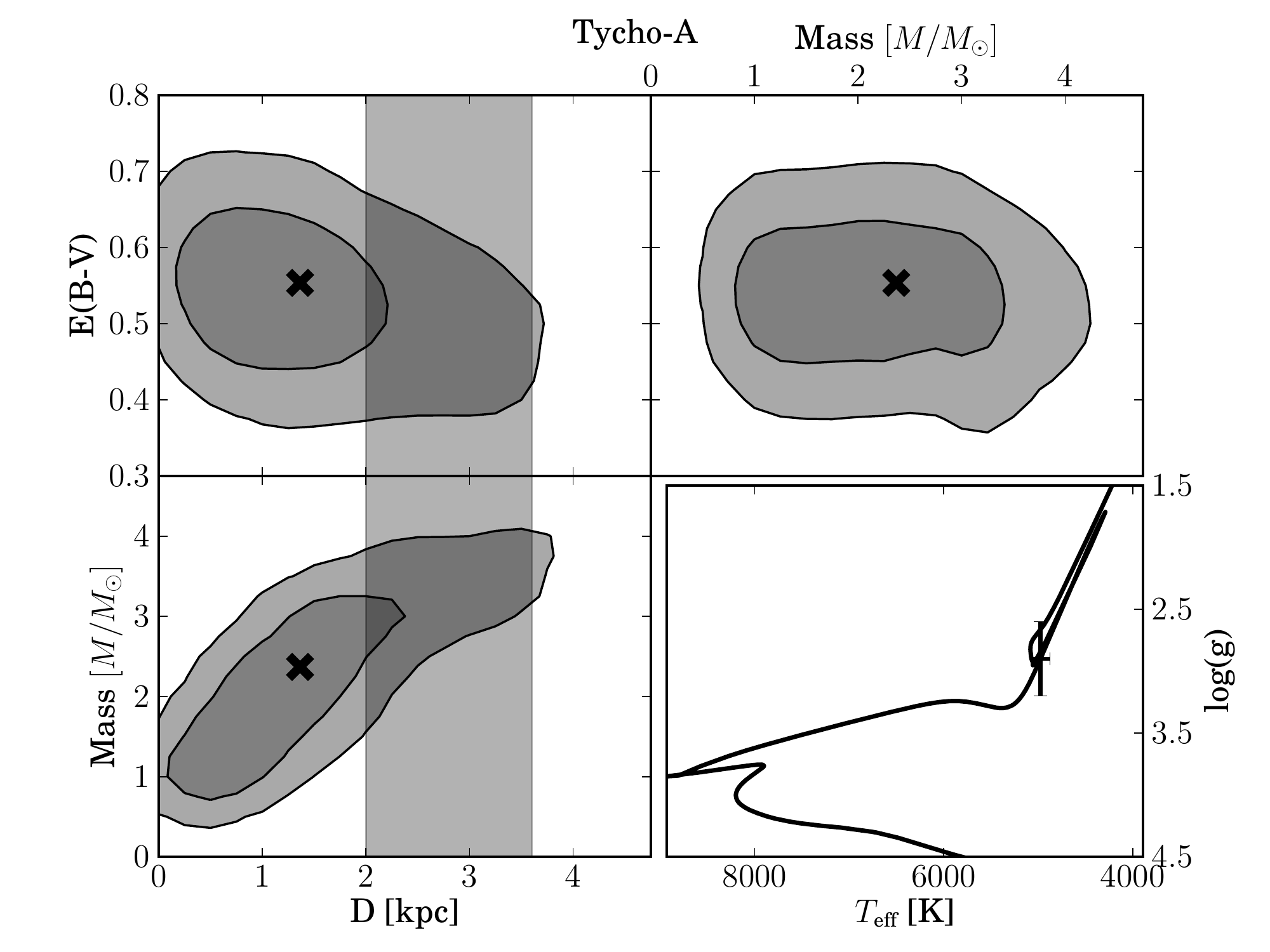} 
   \includegraphics[width=0.5\textwidth]{\plotdir 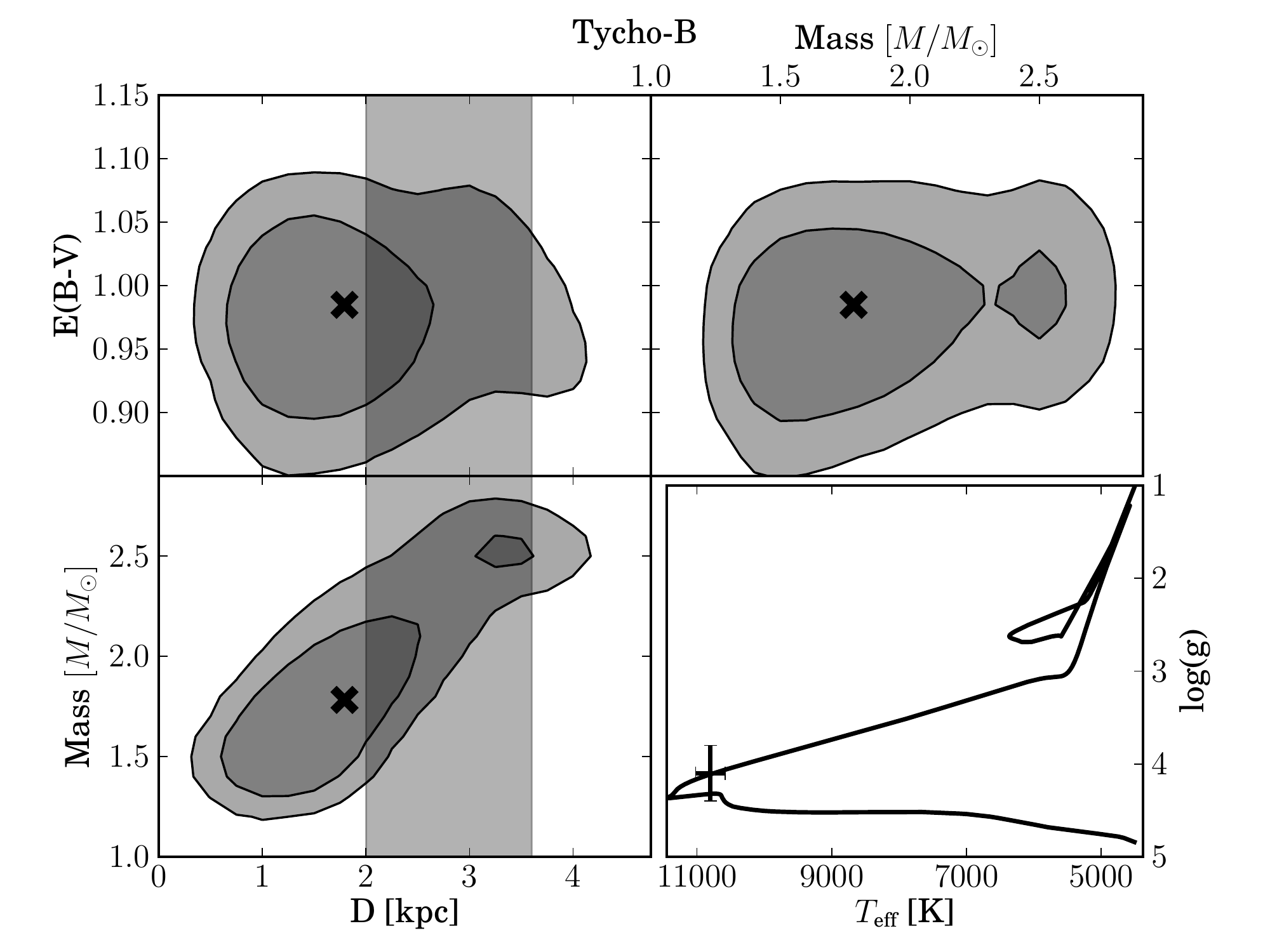} 
   \includegraphics[width=0.5\textwidth]{\plotdir 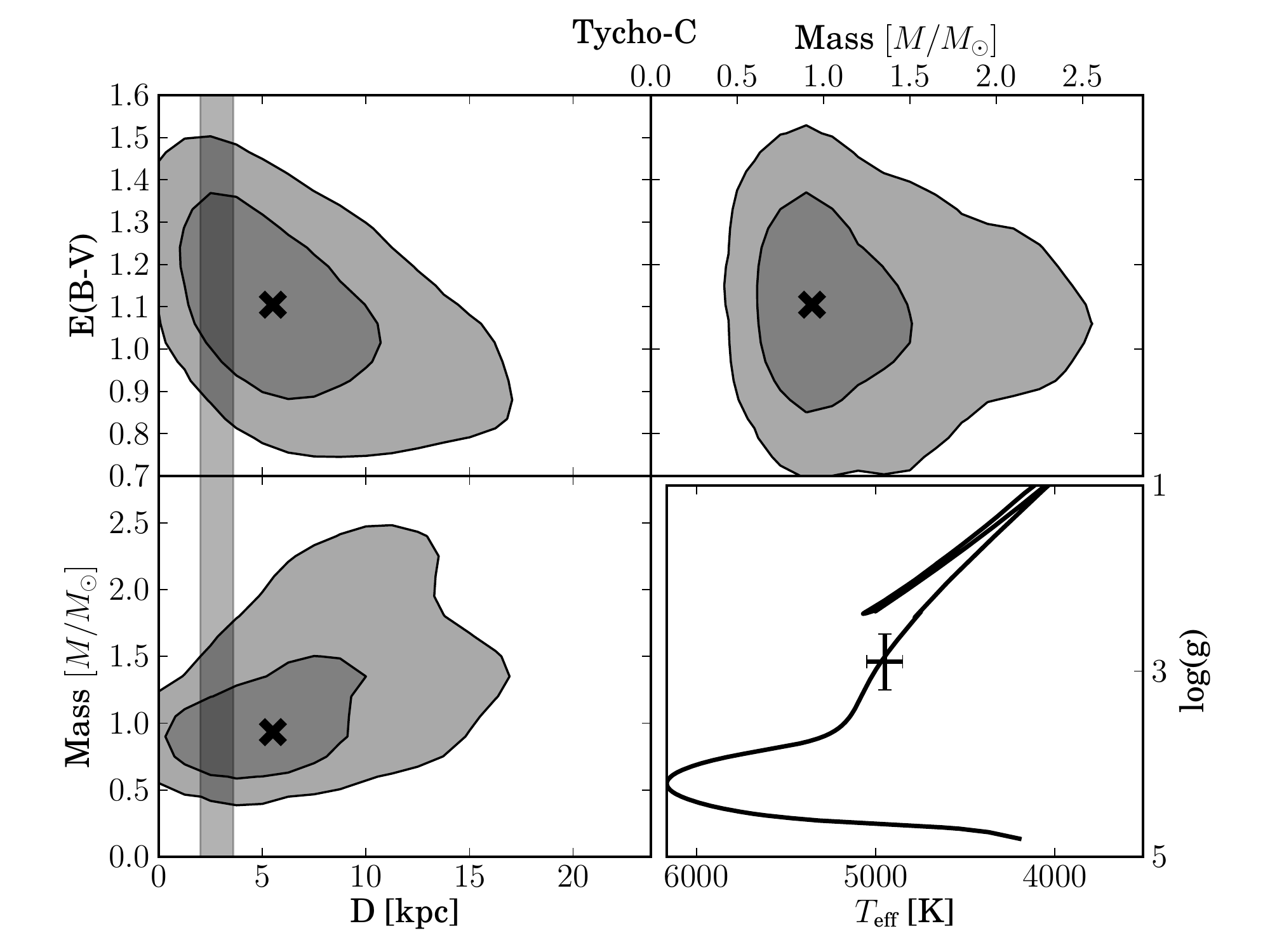} 
   \includegraphics[width=0.5\textwidth]{\plotdir 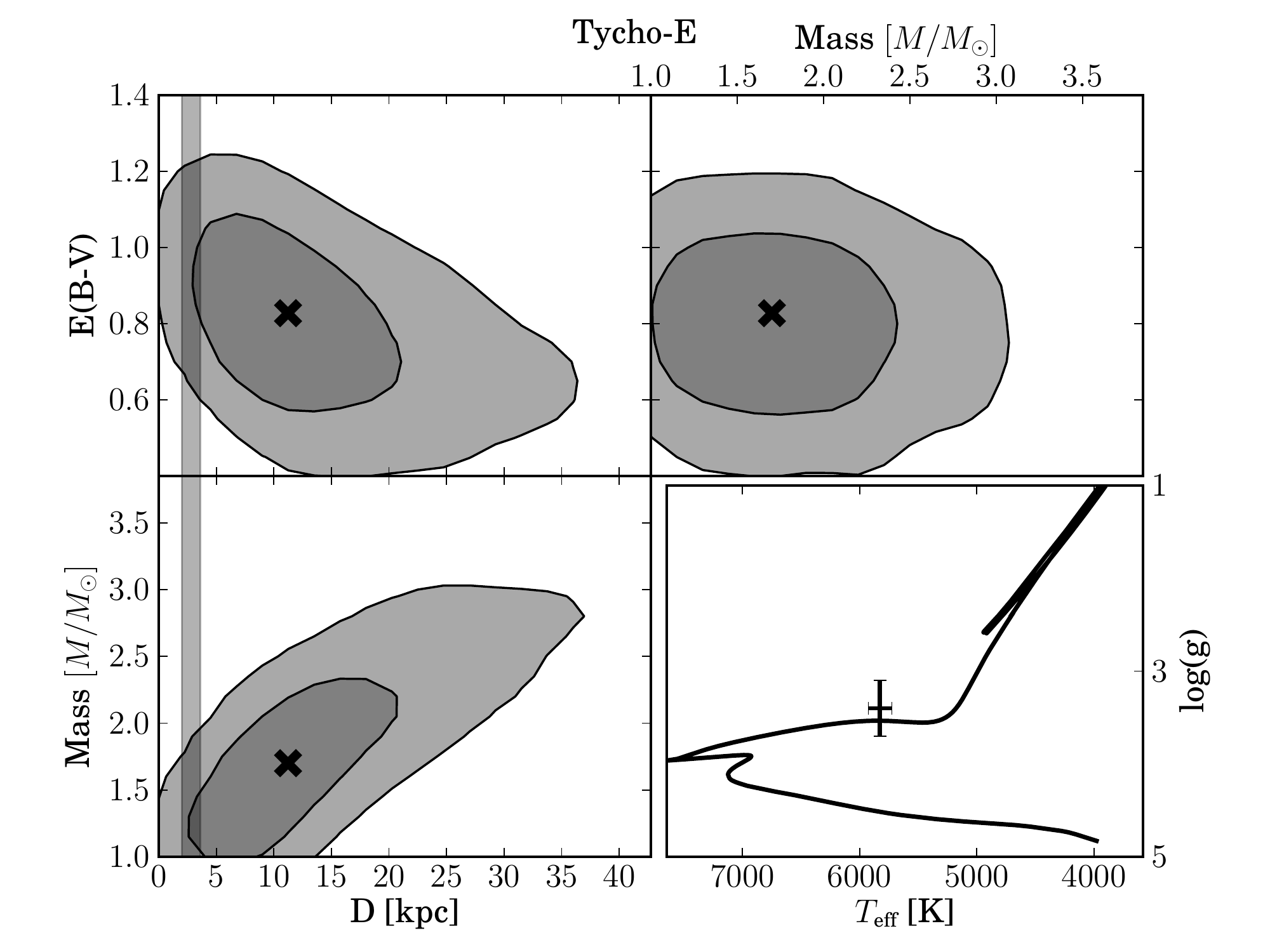} 
   \includegraphics[width=0.5\textwidth]{\plotdir 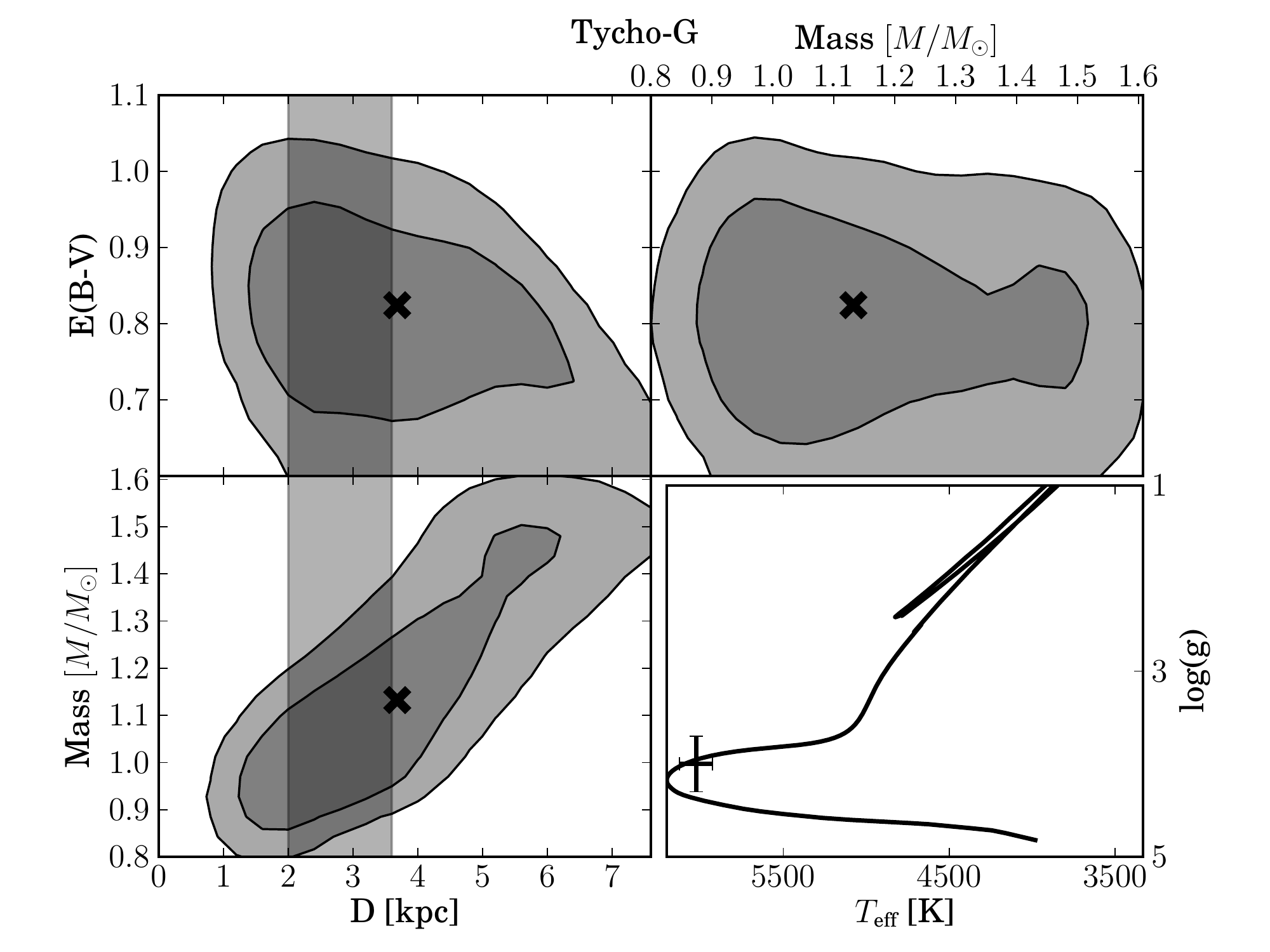} 
   \caption[Distance, extinction, and mass measurements in SN~1572]{The figures show error contours for distance, extinction, and mass of the candidates. In the distance plots we indicate the distance range of SNR~1572 with a gray shade. The lower right shows the optimal isochrone \citep{2004ApJ...612..168P}  for the measured values of $T_{\rm eff}$ and $\log\,{g}$. }
   \label{fig:mc_isochrone}
\end{figure*}

To measure the distance to the candidate stars we used colors and absolute magnitudes from isochrones by \citet{2004ApJ...612..168P}. We used the \gls{migrad} algorithm  \citep{James:1975dr} to find close matches of the measured values to $\teff$--$\log\,g$ isochrones by varying the age of the isochrone.  Subsequently we calculated $E(B-V)$ using the isochrone's color, and we extracted a mass from the isochrone. The results can be seen in Table~\ref{tab:iso_dist}. To estimate the uncertainties in all distances, reddenings, and masses, we employed the Monte-Carlo
method with 10,000 samples of \gls{teff}, \gls{logg}, \gls{feh}, $B$ magnitude, and $V$ magnitude (see Figure  \ref{fig:mc_isochrone}).  Errors included in Table~\ref{tab:iso_dist} are the standard deviations of the Monte-Carlo sample. 
The data show that all stars are compatible with the distance of the remnant. This is not unexpected, as the uncertainties of the measurements in stellar parameters are relatively large.


\begin{deluxetable}{lcccccc}
\tablecaption{Distances, Ages, and Masses of Candidate Stars\label{tab:iso_dist}}
\tablecolumns{7} 
\tablehead{ 
\colhead{Tycho} &
\colhead{Mass}&
\colhead{$\sigma_\textrm{Mass}$} &  
\colhead{Age}&
\colhead{$\sigma_\textrm{Age}$} &
\colhead{$D$}  &
\colhead{$\sigma_D$}\\
\colhead{(Name)} &
\colhead{($M/\msun$)} & 
\colhead{($M/\msun$)} & 
\colhead{(Gyr)} & 
\colhead{(Gyr)}&
\colhead{(\kpc)} &
\colhead{(\kpc)}\\
}

\startdata
\stara & 2.4 & 0.8 &0.7 & 2.3 & 1.4 & 0.8\\
\starb & 1.8 & 0.4 &0.8 & 0.3 & 1.8 & 0.8\\
\starc & 0.9 & 0.4 &10.0 & 3.4 & 5.5 & 3.5\\
\stare & 1.7 & 0.4 &1.4 & 1.1 & 11.2 & 7.5\\
\starg & 1.1 & 0.2 &5.7 & 2.1 & 3.7 & 1.5\\
\enddata
\end{deluxetable}

\section{Discussion}
\label{sec:sn1572_hires:discussion}

In our sample of six stars we find no star that shows characteristics which strongly indicate that it might be the donor star of \sn{1572}{}. On the other hand, it is difficult to absolutely rule out any particular star, if one is able to invoke improbable post-explosion evolutionary scenarios.

Tycho-A is a metal-rich giant, and it seems likely to be a foreground star. Its principal redeeming feature as a donor-star candidate is that it is located in the geometric center of the remnant, and that it has a relatively low surface gravity. \stara\ shows a very low spatial motion, which is consistent with a giant-donor-star scenario, although its lack of rotation is in conflict with a donor-star scenario. 
Taking all measurements into account, we regard \stara\ to be a very weak candidate (although a wind accretion scenario might still work).

Tycho-B's  high temperature, position at the center of the remnant, high rotational velocity, and unusual chemical abundance make it the most unusual candidate in the remnant's center. Despite the {\it a posteriori} unlikely discovery of such a star in the remnant's center, \starb's high rotational velocity coupled with its low spatial velocity seem to be in conflict with any viable donor-star scenario. 
These scenarios predict that the donor star will tidally couple to the white dwarf before explosion, causing the rotation and spatial motion to be correlated post explosion (as discussed in \wek). The large rotation seen in \starb\ should be accompanied by a large spatial motion, which is ruled out by the observations presented here, a problem we are unable to reconcile with \starb\ being the donor star. 
However, Tycho-B does show some unusual abundances, which we will scrutinize in future studies.

Tycho-C consists of two stars which are resolved only in {\it HST} images. It consists of a brighter bluer component ($B = 21.28$, $V = 19.38$, $R = 18.10$ mag; \rl) and a dimmer redder component ($B = 22.91$, $V = 20.53$, $R = 19.23$ mag; \rl). In our analysis we find a consistent solution for the spectrum and infer that this is from the brighter bluer component. 
We find that \starc\ is a metal-poor giant, probably located beyond the remnant. \starc, similarly to \stara, might be compatible with a giant-donor-star scenario. Its lack of rotation and kinematics, however, make it an uncompelling candidate. The only information we have about the dimmer component is the proper motion, which is insignificant with $\mu_\alpha= 0.58 \pm 1.73\,\masyr, \mu_\delta = -0.29 \pm 1.21$\,\masyr. 


Tycho-D is roughly a factor of 10 dimmer than the nearby star \starc\ (separation $\approx 0.6$\arcsec). We could not measure reliable EWs for spectra with this S/N. Visual inspection of the star's spectral features shows it to be consistent with a cool star with low rotation. Its luminosity precludes it from being a relatively slowly rotating giant, and its slow rotation precludes it from being a subgiant or main-sequence donor star. All of this suggests that \stard\ is an uncompelling donor candidate. 

Tycho-E is the most distant star in this set (11.2\,kpc), although large uncertainties in the distance remain. It seems to be similar to \starg\ in temperature, but appears to have a lower surface gravity. It is located 7\arcsec\ from the geometric center, but has no unusual stellar parameters or kinematics. \gh\ have suggested this to be a double-lined binary, but we are unable to confirm this using Fourier cross-correlation techniques.  \citet{2007PASJ...59..811I} have looked at iron absorption lines in stellar spectra made by the remnant and found \stare\ to be unusual. They suggest that a star in the background would show blueshifted and redshifted iron lines, whereas a star inside the remnant would only show blueshifted iron lines, and a foreground star would not show any iron features from the remnant. \citet{2007PASJ...59..811I} claim that \stare\ only shows blueshifted lines, and thus suggest that it is inside the remnant. We believe, however, that \stare\ is located far behind the remnant and suggest that a low column density on the receding side of the remnant could cause a lack of redshifted iron features.  In summary, a lack of rotation, kinematic signatures, and an inconsistent distance make \stare\ a very weak candidate.

Tycho-G is located 30\arcsec\ from the \xray\ center, making it the most remote object from the center in this work (in the plane of the sky; for comparison a distance of 32.6\arcsec\ corresponds to 1000\,\kms\ over 433 yr at the distance of 2.8\,\kpc). This work confirms the radial velocity measured by \gh\ and \wek. Figure \ref{fig:dist_vr} shows the expected distribution of radial velocities from the Besan\c{c}on model of Galactic dynamics. \starg\ lies well within the expected range of \gls{vrad} for stars with its stellar parameters and distance.

\begin{figure*}[ht!] 
   \centering
   \includegraphics[width=\textwidth, trim=0 0 2cm 0, clip]{\plotdir 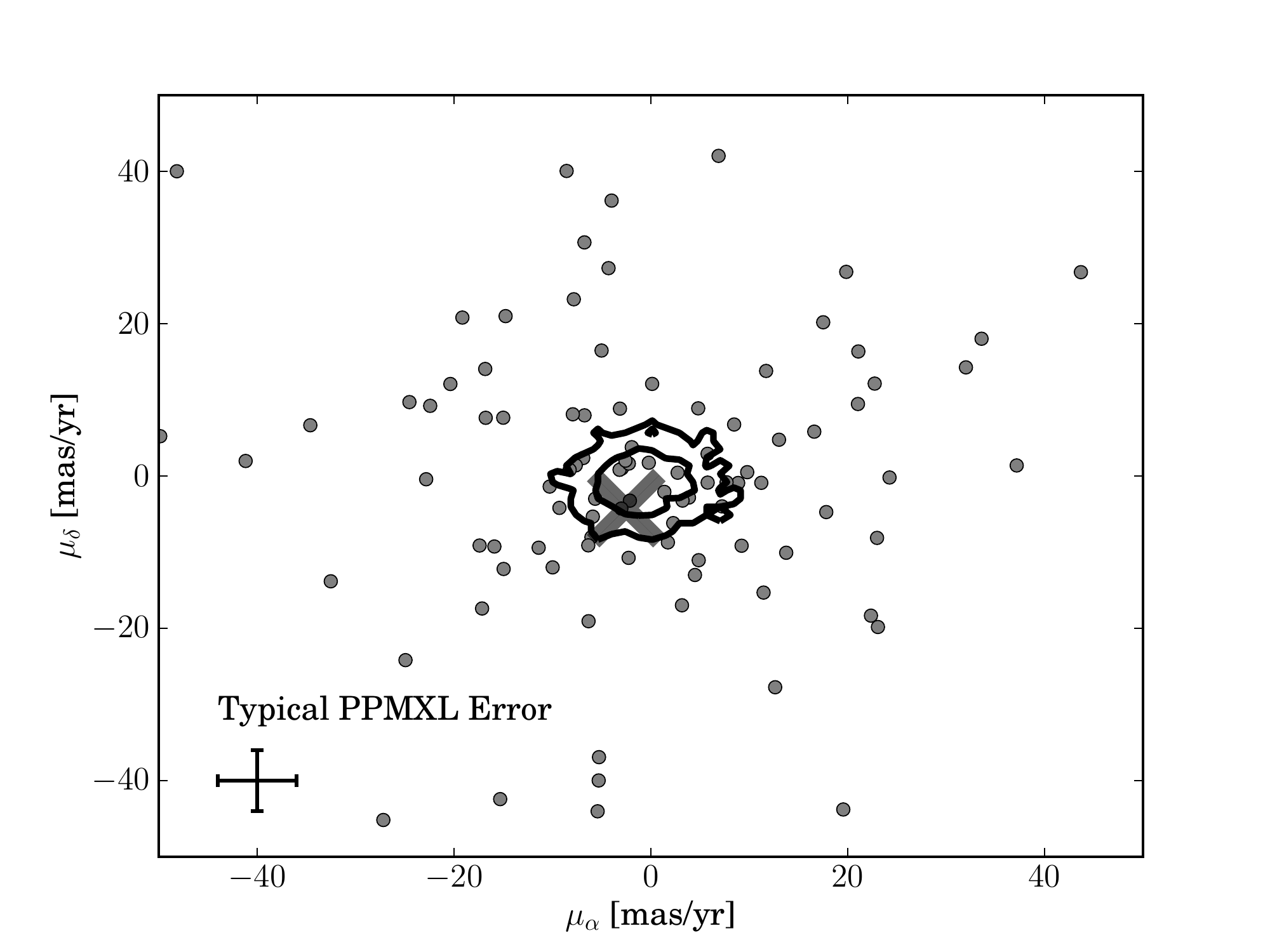} 
   \caption[Comparison between PPMXL catalog and the Besan\c{c}on Model]{The gray points are \glsentryname{ppmxl} stars within $1^\circ$ of \sn{1572}{}. \starg\ has been marked with a gray cross. In addition, we show the distribution for the Besan\c{c}on Model with an area of 1 square degree (a third of the search area of the \gls{ppmxl} sample) around the remnant and a distance between 2.2\,\kpc\ and  5.2\,\kpc\ as the black contours ($1\sigma$, $2\sigma$, and $3\sigma$). We have constrained the model output to only show thick-disk stars with which \starg\ seems to be consistent. }
   \label{fig:sn1572_hires:ppmxl_compare}
\end{figure*}

In addition, this work has analyzed the proper motion of stars around the center of \sn{1572}{}. Figures \ref{fig:propmot_sn1572_hires} and \ref{fig:sn1572_hires:ppmxl_compare} show \starg\ not to be significantly deviant from the distribution of proper motions in the \snr{1572} neighborhood. Figure~\ref{fig:propmot_sn1572_hires} shows \starg\ to be a $2\sigma$ outlier, which implies that there should be about six stars in the {\it HST} sample sharing similar proper-motion features as \starg; thus, its proper motion is by no means a unique trait. To further explore the proper-motion parameter space, we have selected candidates within a $1^\circ$ radius around \sn{1572}{}\ from the proper-motion catalogue \gls{ppmxl} \citep{2010AJ....139.2440R}.  To exclude the many foreground stars we introduced the additional selection criteria $R>16$ mag and $V-R < 1$ mag (for comparison, the Sun has a color of $V-R=1.3$ mag). These selection criteria are meant to exclude foreground stars. We tested this by applying the same selection criteria on the Besan\c{c}on Model, resulting in 95\% of stars more distant than 2\,\kpc. This shows that a high proper motion at great distances is not a unique feature, as there are more stars that share this trait (see Figure \ref{fig:sn1572_hires:ppmxl_compare}). In particular, stars in the thick disk have motions entirely consistent with Tycho-G (see contours in Figure \ref{fig:sn1572_hires:ppmxl_compare}, and Figure 10 in \gh). Finally, the {\it HST} proper-motion measurements are challenging, and it is conceivable that there are systematic errors in our proper-motion measurements which are larger than our reported statistical errors. Such errors would tend to increase the chance of larger-than-actual proper-motion measurements. Taken in total, while \starg\ may have an unusual proper motion, the significance of this motion, even if current measurements are exactly correct, is not exceptional.

As described, the kinematic features of a donor star might easily be lost in the kinematic noise of the Galaxy. \wek\ recommend using post-explosion stellar rotation as an additional possible feature for a donor star. This work suggests that \starg\ has a rotation below the instrumental profile  of 6\,\kms, much less than expected for a donor star \citep[for an estimate, see][]{2009ApJ...701.1665K}. New results by \citet{2012ApJ...750..151P}, however, suggest that only taking tidal coupling into account could overestimate the rotation, and thus \starg's low rotation might still be reconcilable with a donor-star model.

We find \starg\ to be a subgiant/main-sequence star with roughly solar temperature and metallicity.
\gh\ measure a nickel enhancement, which they believe to originate in the contamination from the ejecta. We have conducted a detailed comparison with \gh's measurement in \S~\ref{sec:tychog_comp} and do not find \starg\ to be an outlier as suggested by \gh, but rather consistent with other stars of similar metallicity. In addition, our Li measurement is in agreement with that of \gh\ (see Table~\ref{tab:parvar}). In contrast to the GH09 interpretation, this Li abundance is consistent with that of stars of similar parameters \citep{2010A&A...519A..87B}.
Finally, we have measured the distance to \starg, showing it to be consistent with a background star. In addition, the radial-velocity signature matches that of background stars (see Figure \ref{fig:dist_vr}).

In summary, while \starg\ may have unusual kinematics as indicated by its proper motion, the significance of this motion is not large when compared to a large sample of similar stars in the direction of the Tycho remnant. Furthermore, such a kinematic signature, if it were related to the binary orbital velocity, might predict rotation for \starg\ which we do not observe (modulo the caveats from \wek\ \& \citealt{2012ApJ...750..151P}). Also, we have not  found a reasonable explanation for \starg's large distance from the geometric center, and suggest that \starg\ is unlikely to be related to the Tycho SNR.

\section{Conclusion}
\label{sec:sn1572_hires:conclusion}
This work did not detect an unambiguously identifiable donor-star candidate. Although \starb\ shows some unusual features, there currently remains no convincing explanation for all of its parameters which can be attributed to the donor-star scenario. We believe that our results provide evidence that the Tycho SNR does not have a main-sequence, subgiant, or red giant donor star. Some other possibilities remain. In the spin-down scenario, the companion star can become a helium 
white dwarf from a red giant donor, or a very low mass main-sequence
star from a more massive main-sequence star.  Such a compact companion
can escape detection \citep{2011ApJ...738L...1D, 2011ApJ...730L..34J, 2012ApJ...756L...4H,2012ApJ...744...69H}.  Another scenario is a helium donor, such as the so-called sub-\glsentryname{mchan} explosions discussed by \citet{1995ApJ...452...62L} and \citet{2010ApJ...714L..52S}. These progenitor systems might leave a very faint and fast-moving helium star, or no remnant at all (R.~Pakmor 2012, privat communication). Such a progenitor would probably evade detection, and would likely not leave traces, such as circumstellar interaction with the remnant, or early light-curve anomalies \citep{2010ApJ...708.1025K}. Deep multi-epoch wide-field optical images should catch any such star speeding away from the remnant's center, but such observations have not yet been taken. Finally, a double-degenerate progenitor, in most cases, does not leave a compact remnant, and is consistent with our finding no donor star in \snr{1572}. 

\sn{1006}{}\ and \sn{1604}{}\ (Kepler's \gls*{sn}) are two other \snia\ remnants in the Milky Way. \sn{1006}{}\ is far from the Galactic plane and shows no signs of circumstellar interaction. Kerzendorf et al. (2012b) have studied this remnant and have not found any unusual star that can be explained with a donor-star scenario (consistent with this work). \snr{1604}, while far from the Galactic plane, shows circumstellar interaction with its remnant, and has all the indications of what might be expected from a single-degenerate scenario with an asymptotic giant branch donor \citep{2011arXiv1103.5487C}. Observations of these remnants will better establish if there is a continued pattern to the unusual stars in \snia\ remnant centers, or whether the lack of viable donor stars persists in multiple systems.

\section{Acknowledgements}

B.~P.~Schmidt and W.~E.~Kerzendorf were supported by Schmidt's ARC Laureate Fellowship (FL0992131). A.~Gal-Yam acknowledges support by the Israeli Science Foundation. A.~V.~Filippenko is grateful for the support of the Christopher R. Redlich Fund, the TABASGO Foundation, and NSF grants AST-0908886 and AST-1211916; funding was also provided by NASA grants GO-10098, GO-12469, and AR-12623 from the Space Telescope Science Institute, which is operated by AURA, Inc., under NASA contract NAS 5-26555. R.~J.~Foley was supported by a Clay Fellowship.

We thank Christopher Onken and Jorge Melendez for helpful advice on the HIRES data reduction. We acknowledge useful discussions about differential rotation with Amanda Karakas and also thank Peter Wood for advising us on stellar evolution matters. R\"{u}diger Pakmor provided information on helium-star mergers. We thank Martin Asplund for providing us with Li NLTE corrections.

\bibliographystyle{apj}
\bibliography{sn1572_hires}

\end{document}